\documentclass[pre,preprint]{revtex4}
\usepackage[dvips]{graphicx}

\begin{document}
\title{Reducing the Bias of Causality Measures}
\author{A.~Papana}
 \email{agpapana@gen.auth.gr}
 \affiliation{Department of Mathematical, Physical and Computational Sciences, Faculty of
Engineering, Aristotle University of Thessaloniki, Thessaloniki
54124, Greece}
\author{D.~Kugiumtzis}
 \email{dkugiu@gen.auth.gr}
 \homepage{http://users.auth.gr/dkugiu}
 \affiliation{Department of Mathematical, Physical and Computational Sciences, Faculty of
Engineering, Aristotle University of Thessaloniki, Thessaloniki
54124, Greece}
\author{P.~G.~Larsson}
 \affiliation{Department of Neurology, Oslo University Hospital, Norway}

\begin{abstract}
Measures of the direction and strength of the interdependence
between two time series are evaluated and modified in order to
reduce the bias in the estimation of the measures, so that they
give zero values when there is no causal effect. For this, point
shuffling is employed as used in the frame of surrogate data. This
correction is not specific to a particular measure and it is
implemented here on measures based on state space reconstruction
and information measures. The performance of the causality
measures and their modifications is evaluated on simulated
uncoupled and coupled dynamical systems and for different settings
of embedding dimension, time series length and noise level. The
corrected measures, and particularly the suggested corrected
transfer entropy, turn out to stabilize at the zero level in the
absence of causal effect and detect correctly the direction of
information flow when it is present. The measures are also
evaluated on electroencephalograms (EEG) for the detection of the
information flow in the brain of an epileptic patient. The
performance of the measures on EEG is interpreted, in view of the
results from the simulation study.
\end{abstract}


\maketitle

\section{Introduction}
The interaction or coupling between variables or sub-systems of a
complex dynamical system is a developing area of nonlinear
dynamics and time series analysis \cite{Hlavackova07,Arenas08}.
The detection and characterization of interdependence among
interacting components of complex systems can give information
about their functioning and a better understanding of the system
dynamics. Information flow is a ubiquitous feature of many complex
physical phenomena, such as climatic processes
\cite{Donges09,Smirnov09}, electronic circuits \cite{Bezruchko03},
financial markets \cite{Kwon08}, and the brain system
\cite{Arnhold99,Dhamala08}.

Given a set of time series observations, it is essential to assess
whether they originate from coupled or uncoupled systems, estimate
the hidden causal dependencies between them and detect which
system is the driver and which is the responder. Granger causality
has been the leading approach for a long time for inferring the
direction of interactions, based on the predictability of time
series using linear models \cite{Granger69}. If the prior
knowledge of a time series improves the prediction of another, the
former Granger-causes the latter. Many measures have been
developed based on the concept of Granger causality using
cross-spectra and cross-prediction of linear models
\cite{Baccala01,Winterhalder05}. Granger causality has been
extended to incorporate also nonlinear relationships using
nonlinear models \cite{Chen04,Faes08}, or other model-free
measures that exploit nonlinear properties of dynamical systems,
such as measures based on phase and event synchronization
\cite{Rosenblum01,Smirnov03}, reconstruction of the state spaces
\cite{Arnhold99,Quiroga00,Andrzejak03,Romano07,Chicharro09}, and
information theory
\cite{Schreiber00,Palus01,Marschinski02,Staniek08,Hlavackova07,Vejmelka08}.
The information measures make no assumptions on the system
dynamics as opposed to phase or event synchronization measures
that assume strong oscillatory behavior or distinct event
occurrences, respectively, and the state space methods that
require local dynamics being preserved in neighborhoods of
reconstructed points.

Comparative studies on different causality measures reported in
\cite{Palus07,Smirnov05,Lungarella07,Kreuz07} are not conclusive
and do not point to the same measures, also because different
measures are used in each study. In a recent review and evaluation
of state space, synchronization and information causality
measures, we stressed the need to render the statistical
significance of the causality measures as most measures are biased
and indicate causal effects when they are not present
\cite{Papana08Chaos}. A thorough investigation for the validity
and usefulness of a causality measure should start with a test of
significance, i.e. a measure should not identify coupling (or
interaction) in any direction when it is not present. In
statistical terms, this means that the actual probability of
rejection of the null hypothesis when it is true does not exceed
the nominal significance level, usually set to $0.05$. As in any
statistical test, power is of interest after the correct
significance is established, where power here regards the
sensitivity of the causality measure in detecting interaction and
identifying its direction. Some approaches have been proposed to
render significance of the coupling measures using the concept of
surrogate data testing. The so-called effective transfer entropy
uses a random shuffling of the driving time series
\cite{Marschinski02}. Twin surrogates, generated as shadowing
trajectories of the original trajectories, have recently been
suggested to preserve the original individual dynamics
\cite{Thiel06}. Apparently, the closeness of shadowing in the twin
surrogates determines the level at which the coupling is
destroyed. A different and simple way to generate surrogates is to
time-shift the one of the two time series, as suggested in
\cite{Faes08}.

We propose here a different generation of surrogates and shuffle
randomly the reconstructed points of the driving time series,
rather than the samples as done for the effective transfer
entropy. The random shuffling destroys completely the coupling and
the use of the reconstructed points preserves the individual
system dynamics, perhaps not in the same way as by the twin or
time-shifted surrogates. Instead of making a formal surrogate data
test, we use these surrogates to correct the measure and reduce
the bias. The performance of the measures of mean conditional
probability of recurrence \cite{Romano07}, transfer entropy
\cite{Schreiber00} and symbolic transfer entropy \cite{Staniek08},
as well as the respective surrogate-based corrections is assessed
on multiple realizations of uncoupled and coupled nonlinear
systems (maps and flows) for a range of increasing coupling
strengths. In the numerical simulations, the detection of the
coupling directionality and strength is evaluated at different
settings of dynamics complexity, time series length, noise level
and embedding dimensions for the reconstruction of the two state
spaces. All these factors can be sources of bias in the estimation
of the causality measures.

The structure of the paper is as follows. The directional coupling
measures considered in this study are briefly presented in
Sec.~\ref{sec:CausalityMeasures}, and the proposed corrections of
the measures in Sec.~\ref{sec:Modifications}. The results of the
application of the measures and their corrections on simulated
systems are discussed in Sec.~\ref{sec:Simulation} and on a real
application of EEG recordings from epileptic patients in
Sec.\ref{sec:Application}. Finally, the conclusions are drawn in
Sec.~\ref{sec:Discussion}.

\section{Causality Measures}
\label{sec:CausalityMeasures}
Let $\{x_t\}$ and $\{y_t\}$, $t=1,\ldots,n$, denote two
simultaneously observed time series derived from the dynamical
systems $X$ and $Y$, respectively. We formulate the causality
measures for the causal effect of system $X$ on system $Y$,
denoted as $X \rightarrow Y$. For the opposite direction $Y
\rightarrow X$ the formulation is analogous. Let $m_x$ and $m_y$
be the embedding dimensions and $\tau_x$ and $\tau_y$ the delays
for the state space reconstructions of the two systems,
respectively, giving the reconstructed points
$\textbf{x}_t=(x_t,x_{t-\tau_x},\ldots,x_{t-(m_x-1)\tau_x})'$ and
$\textbf{y}_t=(y_t,y_{t-\tau_y},\ldots,y_{t-(m_y-1)\tau_y})'$,
where $t=1,\ldots,n^{\prime}$ and $n^{\prime}
=n-\max\{(m_x-1)\tau_x,(m_y-1)\tau_y\}$. The steps ahead or time
horizon to address the interaction is denoted by $h$.

\subsection{Mean conditional probability of recurrence}
A state space causality measure based on recurrence quantification
analysis \cite{Eckmann87} has been recently introduced, termed the
mean conditional probability of recurrence \cite{Romano07}. Let
\[
R_{i,j}^X = \Theta(\varepsilon_x-\|\textbf{x}_i-\textbf{x}_j\|), \,\,\,
R_{i,j}^Y = \Theta(\varepsilon_y-\|\textbf{y}_i-\textbf{y}_j\|),  \,\,\,
i,j=1,\ldots,n^{\prime}
\]
be the recurrence matrixes of $X$ and $Y$, respectively, where
$\Theta(\cdot)$ is the Heaviside function counting points with
distance smaller than the predefined distance thresholds
$\varepsilon_x$ and $\varepsilon_y$, respectively. The joint
recurrence matrix of $(X,Y)$ is defined as
\[
J_{i,j}^{X,Y} =
\Theta(\varepsilon_x-\|\textbf{x}_i-\textbf{x}_j\|)
\Theta(\varepsilon_y-\|\textbf{y}_i-\textbf{y}_j\|),
 \,\,\, i,j=1,\ldots,n^{\prime},
\]
i.e. a joint recurrence occurs if the system $X$ recurs in its own
phase space and simultaneously, the system $Y$ recurs also in its
own phase space. The mean conditional probability of recurrence
(MCR) is defined as
\begin{equation}
\mbox{MCR}_{X \rightarrow Y} = {\frac{1}{n^{\prime}}}    %
\sum_{i=1}^{n^{\prime}}{\frac{\sum_{j=1}^{n^{\prime}}    %
{J_{i,j}^{X,Y}}} {\sum_{j=1}^{n^{\prime}}{R_{i,j}^Y}}}.  %
\label{eq:MCR}
\end{equation}
If $X$ drives $Y$, then $\mbox{MCR}_{X \rightarrow Y}
> \mbox{MCR}_{Y \rightarrow X}$. The concept of recurrence has been
used to quantify a weaker form of synchronization, and MCR is an
extension of it that detects the direction of coupling
\cite{Romano07}.

\subsection{Transfer entropy}
Transfer entropy (TE) quantifies the information flow from $X$ to
$Y$ by the amount of information explained in $Y$ at one step
ahead (or generally $h$ steps ahead) by the state of $X$,
accounting for the concurrent state of $Y$ \cite{Schreiber00}. The
concept of TE extends the Shannon entropy to transition
probabilities and quantifies how the conditioning on $X$ change
the transition probabilities of $Y$. Using the reconstructed
points for $X$ and $Y$ as given above, TE is defined as
\begin{equation}
\mbox{TE}_{X\rightarrow Y} = \sum p(y_{t+h},\textbf{x}_t,\textbf{y}_t)     %
\log{\frac{p(y_{t+h}|\textbf{x}_t,\textbf{y}_t)}{p(y_{t+h}|\textbf{y}_t)}},%
\label{eq:TE}
\end{equation}
where $p(y_{t+h},\textbf{x}_t,\textbf{y}_t)$,
$p(y_{t+h}|\textbf{x}_t,\textbf{y}_t)$, and
$p(y_{t+h}|\textbf{y}_t)$ are the joint and conditional
probability mass functions for a proper binning. The time horizon
$h$ is introduced here instead of the time step one that was
originally used in the definition of TE. TE can also be defined in
terms of entropies as
\begin{equation}
\mbox{TE}_{X \rightarrow Y} =                                       %
H(\textbf{x}_t,\textbf{y}_t) - H(y_{t+h},\textbf{x}_t,\textbf{y}_t) %
+ H(y_{t+h},\textbf{y}_t) - H(\textbf{y}_t).                        %
\label{eq:TEentropies}
\end{equation}

Instead of binning, we define TE in terms of correlation sums as
follows. Let $X$ be a continuous, possibly vector-valued, random
variable. For a fixed small $r$, the entropy of a variable $X$ can
be estimated as $H(X) \simeq \ln{C(\textbf{x}_t)} + m \ln{r}$
\cite{Manzan02}, where $C(\textbf{x}_t)$ is the correlation sum
for the vectors $\textbf{x}_t$ with embedding dimension $m$
($C(\textbf{x}_t)$ is an estimate of the probability of points
being closer than $r$). The standardized Euclidian norm, i.e. the
Euclidean distance divided by the square root of the embedding
dimension, is used for the calculation of the correlation sum. Let
us denote the correlation sums of the vectors
$[y_{t+h},\textbf{x}_t,\textbf{y}_t]$, $\textbf{y}_t$,
$[\textbf{x}_t,\textbf{y}_t]$ and $[y_{t+h},\textbf{y}_t]$ as
$C(y_{t+h},\textbf{x}_t,\textbf{y}_t)$, $C(\textbf{y}_t)$,
$C(\textbf{x}_t,\textbf{y}_t)$ and $C(y_{t+h},\textbf{y}_t)$,
respectively. Then, TE is defined as
\begin{equation}
\mbox{TE}_{X \rightarrow Y} =                                   %
\log{\frac{C(y_{t+h},\textbf{x}_t,\textbf{y}_t)C(\textbf{y}_t)} %
{C(\textbf{x}_t,\textbf{y}_t)C(y_{t+h},\textbf{y}_t)}}.         %
\label{eq:TECorSum}
\end{equation}

\subsection{Symbolic transfer entropy}
Symbolic transfer entropy (STE) is the transfer entropy defined on
rank-points formed by the reconstructed vectors of $X$ and $Y$
\cite{Staniek08}. Thus, for each vector $\textbf{y}_t$, the ranks
of its components assign a rank-point $\hat{\textbf{y}}_t = [r_1,
r_2, \ldots, r_{m_y}]$, where $r_j \in \{1,2,\ldots,m_y\}$ for
$j=1, \ldots, m_y$. Following this sample-point to rank-point
conversion, the sample $y_{t+h}$ in Eq.(\ref{eq:TE}) is taken as a
rank point at time $t+h$, $\hat{\textbf{y}}_{t+h}$, and STE is
defined as
\begin{equation}
\mbox{STE}_{X\rightarrow Y} =
H(\hat{\textbf{x}}_t,\hat{\textbf{y}}_t) -                            %
H(\hat{\textbf{y}}_{t+h},\hat{\textbf{x}}_t,\hat{\textbf{y}}_t) +     %
H(\hat{\textbf{y}}_{t+h},\hat{\textbf{y}}_t) - H(\hat{\textbf{y}}_t), %
\label{eq:STE}
\end{equation}
where the entropies are defined on the rank-points.

\subsection{Effective transfer entropy}
A modification of TE, called effective transfer entropy (ETE),
was defined in \cite{Marschinski02} as the difference of TE computed
on the original bivariate time series and the TE computed on a
surrogate bivariate time series, where the driving time series $X$
is randomly shuffled
\[
\mbox{ETE}_{X \rightarrow Y} = \mbox{TE}_{X \rightarrow Y} - %
\mbox{TE}_{X_{\mbox{\small shuffled}} \rightarrow Y}.        %
\]
The use of a randomly shuffled surrogate aims at setting a
significance threshold in the estimation of TE. The approach of
ETE can be used for the estimation of any other causal measure.
Here, for the estimation of ETE, instead of one random permutation
a number of $M$ random permutations of the driving time series $X$
are considered and therefore $\mbox{TE}_{X_{\mbox{\small
shuffled}} \rightarrow Y}$ in the definition of ETE is replaced by
the mean of the corresponding $M$ values
\begin{equation}
\mbox{ETE}_{X \rightarrow Y} = \mbox{TE}_{X \rightarrow Y} -
\frac{1}{M} \sum_{l=1}^M \mbox{TE}_{X_{l,\mbox{\small shuffled}}
\rightarrow Y},
 \label{eq:ETE}
\end{equation}
where $l=1,\ldots,M$. In the same way, we define effective STE,
denoted ESTE.

\subsection{Relationship of MCR and TE}
MCR defines in a rather direct way the conditional probability of
close points in $Y$ given they are close in $X$, whereas most
state space methods, such as nonlinear interdependence measures
\cite{Arnhold99,Quiroga00,Andrzejak03,Chicharro09}, attempt to
approximate the conditional probability indirectly through analogy
in distances. To this respect, the MCR method is closer to
information measures, such as TE. However, TE involves also
transition probabilities that can give additional information
about the effect of the driving system on the future of the
response system.

In \cite{Romano07}, it is stressed that MCR needs smaller number
of data points than TE. This is true if binning estimators are
used for the probability functions in TE, but for the estimator
considered here using correlation sums, the data requirements are
the same, as the stability of the estimation in both MCR and TE
relies on having good statistics of points in the neighborhoods
for a given distance. Certainly, this holds when the distance $r$
in TE and the distances $\epsilon_x$ and $\epsilon_y$ in MCR are
at the same level. It is also mentioned in \cite{Romano07} that
TE, but not MCR, may give values larger than zero for both
directions when the coupling is purely unidirectional. As we show
below, this bias is not specific to a measure and should be
attributed to other factors, such as the system complexity and the
length of the time series. The fact that MCR did not exhibit this
bias in the results in \cite{Romano07} may be due to the
optimization of the values of the thresholds $\epsilon_x$ and
$\epsilon_y$, so that for no coupling both averages of the
estimated probabilities of recurrences $p(\textbf{x}_i)$ and
$p(\textbf{y}_i)$ are equal to 0.01. In our simulations we do not
optimize $\epsilon_x$ and $\epsilon_y$ but use a fixed
$\epsilon_x=\epsilon_y = r$ and have the time series standarized,
i.e. the same distance threshold is used in the computation of
both MCR and TE in all simulations.

\section{Modifications of Causality Measures}
\label{sec:Modifications}
A main drawback of all causality measures considered in this study
is that they do not provide stable and consistent results,
particularly for weak coupling structures and noisy time series.
The measures have bias that may be different in each direction,
depending also on the dynamics of each system, the time series
length and the state space reconstruction. The existence of bias
and spurious detection of causal effects has been previously
reported for different causality measures
\cite{Palus07,Lungarella07,Hlavackova07,Papana08Chaos}. When there
is no causal effect the positive bias may be misinterpreted as
weak coupling.

A possible solution to this problem is provided by reducing the
bias of the measure using surrogate data. Surrogate time series
can be used to rule out spurious conclusions about the existence
and the direction of coupling. When testing the null hypothesis
that the two time series are uncoupled, the bivariate surrogates
should replicate the dynamics of each system and be independent to
each other. In this way, the bias due to the individual system
dynamics and state space reconstruction is preserved in the
surrogates. Here, we do not apply a formal surrogate data test,
but we use the surrogate values to correct for the bias of the
coupling measure, as shown below.

The approach in ETE attempts to generate surrogates for this
purpose by randomizing the temporal structure of the driving time
series, so that if the systems are coupled, cause and effect are
lost. However, by randomly shuffling a time series, its
self-dynamical structure is destroyed as well. We present below
correction to the measures MCR, TE and STE, based on the frame of
surrogates, which are extracted by randomly shuffling the
reconstructed vectors of the driving time series in order to
preserve the dynamical properties of each system.

\subsection{Corrected MCR}
The corrected MCR (CMCR) defined below aims at reducing the bias
of MCR in case of no causal effects. Recall that the ones at each
line $i$ of the joint recurrence matrix $J_{i,j}^{X,Y}$,
$i,j=1,\ldots,n^{\prime}$, correspond to the matched time indices
of the neighboring points to $\textbf{x}_i$ in system $X$ and the
neighboring points to $\textbf{y}_i$ in system $Y$, and the number
of matches determines the strength of coupling. Thus by random
shuffling the lines of matrix $R_{i,j}^X$, $i,j =
1,\ldots,n^{\prime}$, we destroy this match, as for each
$\textbf{y}_i$, the neighbors for the $X$ system do not regard any
more $\textbf{x}_i$ but another randomly chosen point. Repeating
this random shuffling $M$ times, we get $M$ new matrices
$Rs_{l,{i,j}}^X$, $l=1,\ldots,M$, and $M$ new joint recurrence
matrices $Js_{l,{i,j}}^{X,Y}$. This allows us to take the average
number of common neighbors at each time index $i$ over the $M$
realizations that regards the scenario of no coupling. The
'surrogate' MCR is then defined as
\[
\mbox{MCRs}_{X \rightarrow Y} = \frac{1}{n^{\prime}}    %
\sum_{i=1}^{n^{\prime}} \frac{\frac{1}{M} \sum_{l=1}^M  %
\sum_{j=1}^{n^{\prime}}{Js_{l,{i,j}}^{X,Y}}}            %
{\sum_{j=1}^{n^{\prime}}{R_{i,j}^Y}}.%
\]
and CMCR is
\begin{equation}
\mbox{CMCR}_{X \rightarrow Y} =
\mbox{MCR}_{X \rightarrow Y}-\mbox{MCRs}_{X \rightarrow Y}.%
\label{eq:CMCR}
\end{equation}

\subsection{Corrected TE and STE}
For the estimation of the corrected TE (CTE), the same idea is
implemented and we assume again $M$ random shufflings of the
points of the $X$ system. Thus in the estimation of TE in
Eq.(\ref{eq:TECorSum}), the terms of the correlation sums
$C(y_{t+h},\textbf{x}_t,\textbf{y}_t)$ and
$C(\textbf{x}_t,\textbf{y}_t)$ are replaced by the corresponding
mean values of the correlation sums estimated on the point
shuffled surrogates, given as
\[
C_s(y_{t+h},\textbf{x}_t,\textbf{y}_t) = \frac{1}{M}
\sum_{l=1}^{M} C(y_{t+h},\textbf{x}_{t_{l}},\textbf{y}_t)
\]
and
\[
C_s(\textbf{x}_t,\textbf{y}_t) = \frac{1}{M} \sum_{l=1}^{M}
C(\textbf{x}_{t_l},\textbf{y}_t),
\]
where $t_l$ denotes a random time index and $\textbf{x}_{t_l}$ is
the point in system $X$ at the $l$-th replication for the time
index $t$. Then, the 'surrogate' TE value is estimated as
\[
\mbox{TEs}_{X \rightarrow Y} =                                   %
\log{\frac{C_s(y_{t+h},\textbf{x}_t,\textbf{y}_t)C(\textbf{y}_t)}%
{C_s(\textbf{x}_t,\textbf{y}_t)C(y_{t+h},\textbf{y}_t)}}
\]
and CTE is defined as
\begin{equation}
\mbox{CTE}_{X \rightarrow Y} =                             %
\mbox{TE}_{X \rightarrow Y} - \mbox{TEs}_{X \rightarrow Y}. %
\label{eq:CTE}
\end{equation}
Note that instead of taking the average of $M$ 'surrogate' TE
values as in Eq.(\ref{eq:ETE}) for ETE, a single 'surrogate' TE
value is extracted by taking the average at each term of the
expression of TE. The difference is actually in taking the average
after or before the logarithm of each correlation sum. The former
gives more variable estimates of TE on the surrogates as for small
values of the correlation sums we obtain large negative
logarithms. Thus by taking the mean of the correlation sums over
all surrogates we stabilize the correlation sum to the most
representative value expected if the systems were to be uncoupled.
In turn this gives more stable estimation of the mean entropy
terms and subsequently the mean transfer entropy for the
surrogates.

Substituting in Eq.(\ref{eq:CTE}) the expression for the original
and surrogate TE, the terms $H(\textbf{y}_t)$ and
$H(y_{t+h},\textbf{y}_t)$ cancel out and we get
\[
\mbox{CTE}_{X \rightarrow Y} =
\log{\frac{C(y_{t+h},\textbf{x}_t,\textbf{y}_t)C_s(\textbf{x}_t,\textbf{y}_t)}
{C_s(y_{t+h},\textbf{x}_t,\textbf{y}_t)C(\textbf{x}_t,\textbf{y}_t)}}.
\]
This measure should be zero when $X$ does not have any effect on
$Y$. However, other sources of bias may still cause deviations
from zero even in the lack of causal effect and this will be
tested below through simulations.

Corrected STE (CSTE) is defined analogously to CTE, and the
expression of CSTE in terms of entropies is
\begin{equation}
\mbox{CSTE}_{X \rightarrow Y} =
H(\hat{\textbf{x}}_t,\hat{\textbf{y}}_t) -
H(\hat{y}_{t+h},\hat{\textbf{x}}_t,\hat{\textbf{y}}_t) -
H_s(\hat{\textbf{x}}_t,\hat{\textbf{y}}_t) +
H_s(\hat{y}_{t+h},\hat{\textbf{x}}_t,\hat{\textbf{y}}_t),
\end{equation}
where
\begin{equation}
H_s(\hat{\textbf{x}}_t,\hat{\textbf{y}}_t) = \frac{1}{M}
\sum_{l=1}^M H(\hat{\textbf{x}}_{t_l},\hat{\textbf{y}}_t)
\end{equation}
and
\begin{equation}
H_s(\hat{y}_{t+h},\hat{\textbf{x}}_t,\hat{\textbf{y}}_t) =
\frac{1}{M} \sum_{l=1}^M
H(\hat{y}_{t+h},\hat{\textbf{x}}_{t_l},\hat{\textbf{y}}_t).
\end{equation}

\section{Evaluation of Causality Measures on Simulated Systems}
\label{sec:Simulation}
\subsection{Simulation Setup}
Measures of directional coupling are computed on 100 realizations
of the following unidirectionally coupled systems, for increasing
coupling strengths and for both directions $X \rightarrow Y$ and
$Y \rightarrow X$.
\begin{itemize}
\item Two unidirectionally coupled Henon maps \\
$x_{t+1} = 1.4  - x^{2}_t + 0.3x_{t-1}$ \\
$y_{t+1} = 1.4 - cx_t y_t - (1-c)y^{2}_t + 0.3y_{t-1}$ \\
with coupling strengths $c=0,0.05,0.1,0.2,0.3,0.4,0.5,0.6,0.7$,
whereas the onset to identical synchronization occurs at
approximately $c=0.65$ \cite{Quiroga00,Rosenblum01}.
\item Two unidirectionally coupled Mackey-Glass systems
\cite{Senthilkumar08}
\begin{eqnarray}
\frac{\mbox{d}x}{\mbox{d}t} & = &
\frac{0.2x_{t-\Delta_x}}{1+x_{t-\Delta_x}^{10}}-0.1x_t \nonumber \\ %
\frac{\mbox{d}y}{\mbox{d}t} & = &
\frac{0.2y_{t-\Delta_y}}{1+y_{t-\Delta_y}^{10}}+ c
\frac{0.2x_{t-\Delta_x}}{1+x_{t-\Delta_x}^{10}}-0.1y_t
\label{eq:UnidCoupledMackeyGlass}
\end{eqnarray}
for $\Delta_x$ and $\Delta_y$ taking the values 17, 30, 100 (all 9
combinations for the driving and response system) and with
coupling strengths $c=$ 0, 0.05, 0.1, 0.15, 0.2, 0.3, 0.4 and 0.5.
The choice of the different $\Delta_x$ and $\Delta_y$ aims at
investigating the performance of the measures on systems with the
same and different complexity. For the three delay parameters the
Mackey-Glass system is chaotic with correlation dimension about 2
for $\Delta_x=17$, 3 for $\Delta_x=30$ and 7 for $\Delta_x=100$
\cite{Grassberger83}. The integration was done with the function
{\tt dde23} of the {\tt MATLAB} software and the time series were
obtained with sampling time 4 sec.
\item A coupled nonlinear stochastic system (see
\cite{Gourevitch06}) \\
$x_t = 3.4 x_{t-1} (1-x_{t-1}^2) e^{-x_{t-1}^2} + 0.4 {e_1}_t$ \\  %
$y_t = 3.4 y_{t-1} (1-y_{t-1}^2) e^{-y_{t-1}^2}  + 0.5 x_{t-1} y_{t-1} + 0.4 {e_2}_t$  \\ %
$z_t = 3.4 z_{t-1} (1-z_{t-1}^2) e^{-z_{t-1}^2} + 0.3 y_{t-1} + 0.5 x_{t-1}^2 + 0.4 {e_3}_t$ \\ %
where ${e_1}_t$, ${e_2}_t$ and ${e_3}_t$ are standard white
Gaussian noise processes. We note that the correct directed causal
effects are $X \rightarrow Y$, $Y \rightarrow Z$ and $X
\rightarrow Z$.
\end{itemize}

The time series lengths for the coupled Henon maps are
$n=512,1024,2048$ and for the Mackey-Glass $n=2048$. Gaussian
white noise is also added to these time series with standard
deviation $5\%$ and $20\%$ of the standard deviation of the time
series. Further, we investigate the dependence of the measures on
the state space reconstruction of the two systems. For the coupled
Henon system, the embedding dimensions vary as $m_x=1,\ldots,5$
and $m_y=1,\ldots,5$. For the coupled Mackey-Glass systems, $m_x$
and $m_y$ vary from 1 and up to 10, depending on the delays of the
coupled systems. For symbolic information measures, the embedding
dimensions cannot be set equal to one as there will be no
different symbolic patterns.

In order to obtain quantitative summary results for the
performance of the measures, t-tests for means are conducted for
the following three null hypotheses H$_0$:
\begin{itemize}                                              %
\item H$_0^1$: The mean of the measure in the direction      %
               $X \rightarrow Y$ is zero.                    %
\item H$_0^2$: The mean of the measure for the direction     %
               $Y \rightarrow X$ is zero.                    %
\item H$_0^3$: The means of the measures in the two          %
               directions are equal.                         %
\end{itemize}                                                %
We assume that the distribution of the measure in both directions
formed from its values in 100 realizations is normal and for
H$_0^3$ that they have the same variance, which both seem to be
statistically satisfied (as resulted from the Kolmogorov-Smirnov
test for normality and the Fisher test for equal variances applied
to some of the realizations).

Using as samples the measure values from 100 realizations for each
case, the performance of each measure is quantified in terms of
the rejection or not of each of the three H$_0$ at the
significance level $\alpha=0.05$, giving a score zero if H$_0$ is
not rejected and one if it is rejected. So the total score for all
three H$_0$ ranges from 0 to 3. There are two settings of interest
for the coupling of $X$ and $Y$: no coupling that regards the
significance of the measure, for which the best total score is 0,
and the presence of unidirectional coupling that regards the
discriminating power of the measure, for which the best total
score is 2, meaning rejection of H$_0^1$ and H$_0^3$ but not
H$_0^2$ (the latter yielding the direction of no causal effect).
Note that score 2 can also be obtained if in both directions a
measure is significantly different from zero but at the same
level. Therefore, we will explicitly name the setting for each
H$_0$ when there is ambiguity from the total score.

\subsection{Results on the unidirectionally coupled Henon maps}
\begin{figure} [h!]
\centerline{\hbox{
\includegraphics[height=5.0cm,keepaspectratio]{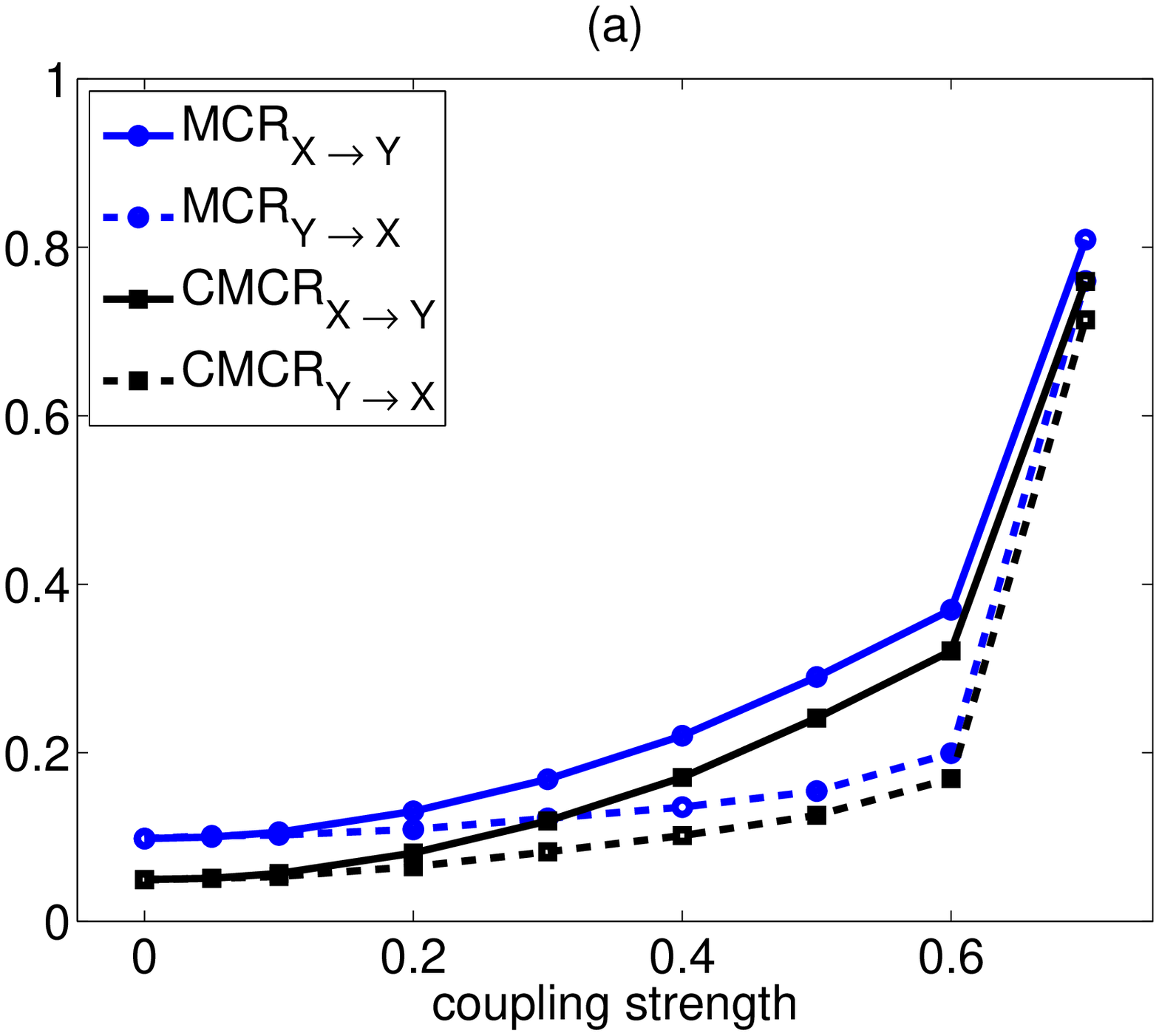}
\includegraphics[height=5.0cm,keepaspectratio]{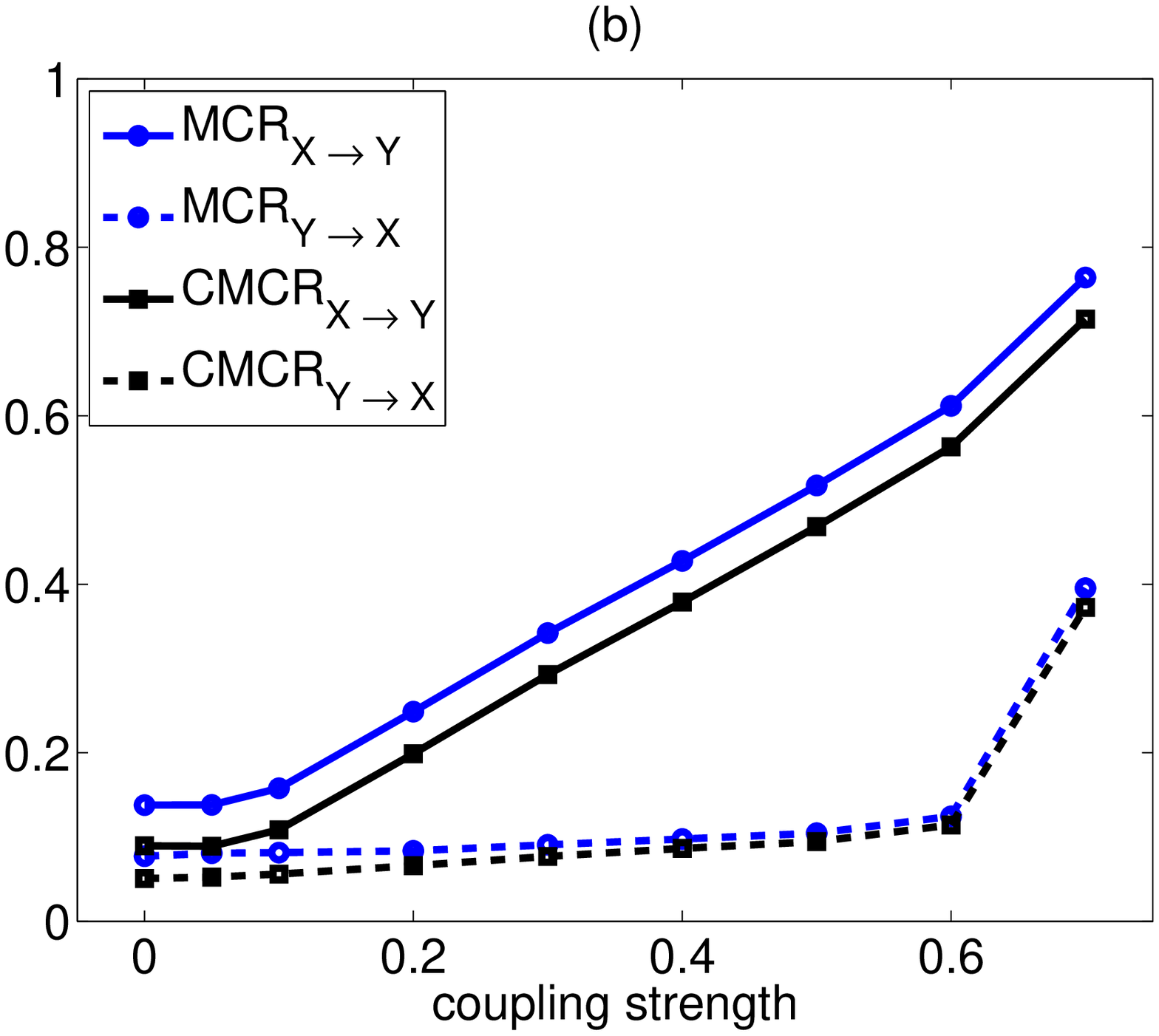}
\includegraphics[height=5.0cm,keepaspectratio]{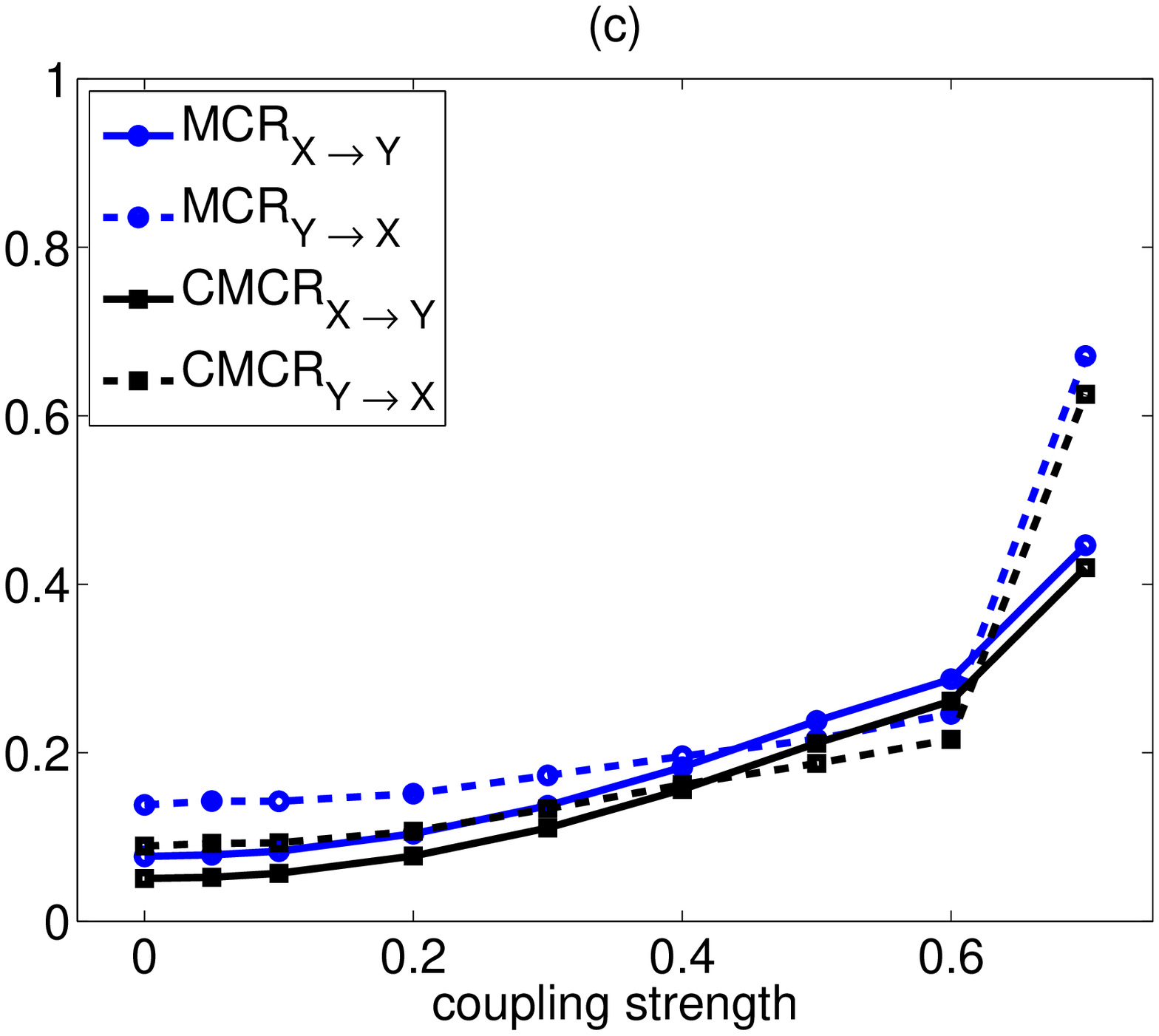}
}} \caption{(a) Mean estimated values of MCR and CMCR for both
directions from $100$ realizations of the noise-free
unidirectionally coupled Henon map, where $m_x=2$, $m_y=2$ and
$n=512$. In (b) and (c) as in (a), but for $m_x=2,m_y=4$ and
$m_x=4,m_y=2$, respectively.} \label{fig:henon0MCR}
\end{figure}

MCR and CMCR measures are significantly affected by the embedding
dimension, time series length and noise level. There are two
important problems with MCR: first, it increases also in the
opposite direction (with no causal effect) with the increase of
the coupling strength, which may erroneously be interpreted as
bidirectional coupling, and second, it is positively biased in the
uncoupled case ($c=0$), especially for short time series. By
increasing the time series length and the embedding dimensions,
MCR and CMCR both decrease. The corrected measure CMCR obtains
always smaller values than MCR in both directions and it is closer
to zero for $c=0$, particularly for small time series and small
embedding dimensions (see Fig.\ref{fig:henon0MCR}).
Both MCR and CMCR maintain a larger increase in the correct
direction of interaction with the coupling strength when
$m_x=m_y$, amplify the difference in the two directions when $m_x
< m_y$ and decrease this difference or even tend to suggest more
interaction in the opposite direction when $m_x > m_y$. Addition
of small noise level ($5\%$) does not substantially affect MCR and
CMCR, but $20\%$ noise level lowers MCR and CMCR toward the zero
level for both directions. In this case, increase of the embedding
dimension regains the correct signature of coupling, but adds
positive bias at a level that is clearly seen for $c=0$.

\begin{figure} [h!]
\centerline{\hbox{
\includegraphics[height=5.0cm,keepaspectratio]{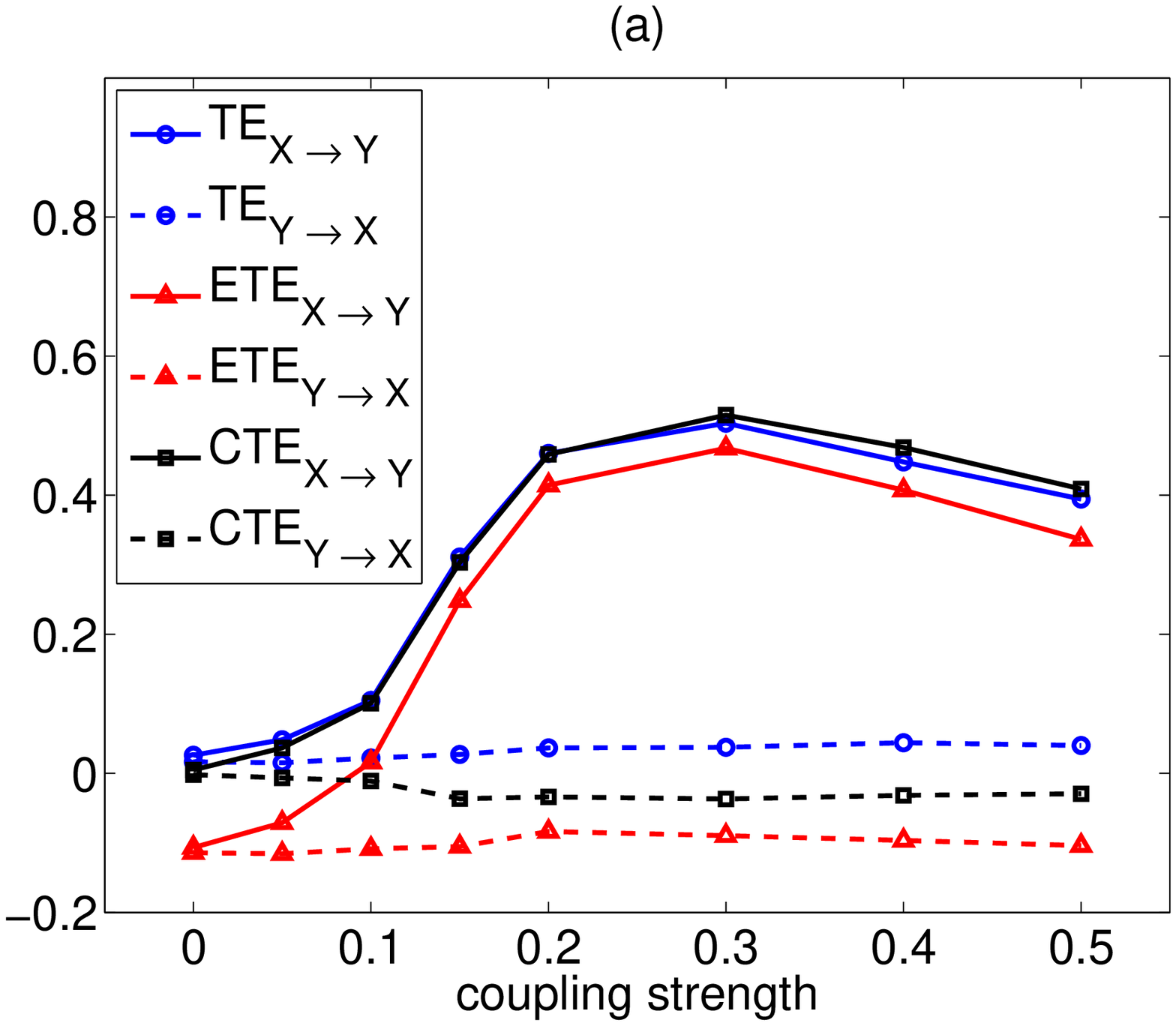}
\includegraphics[height=5.0cm,keepaspectratio]{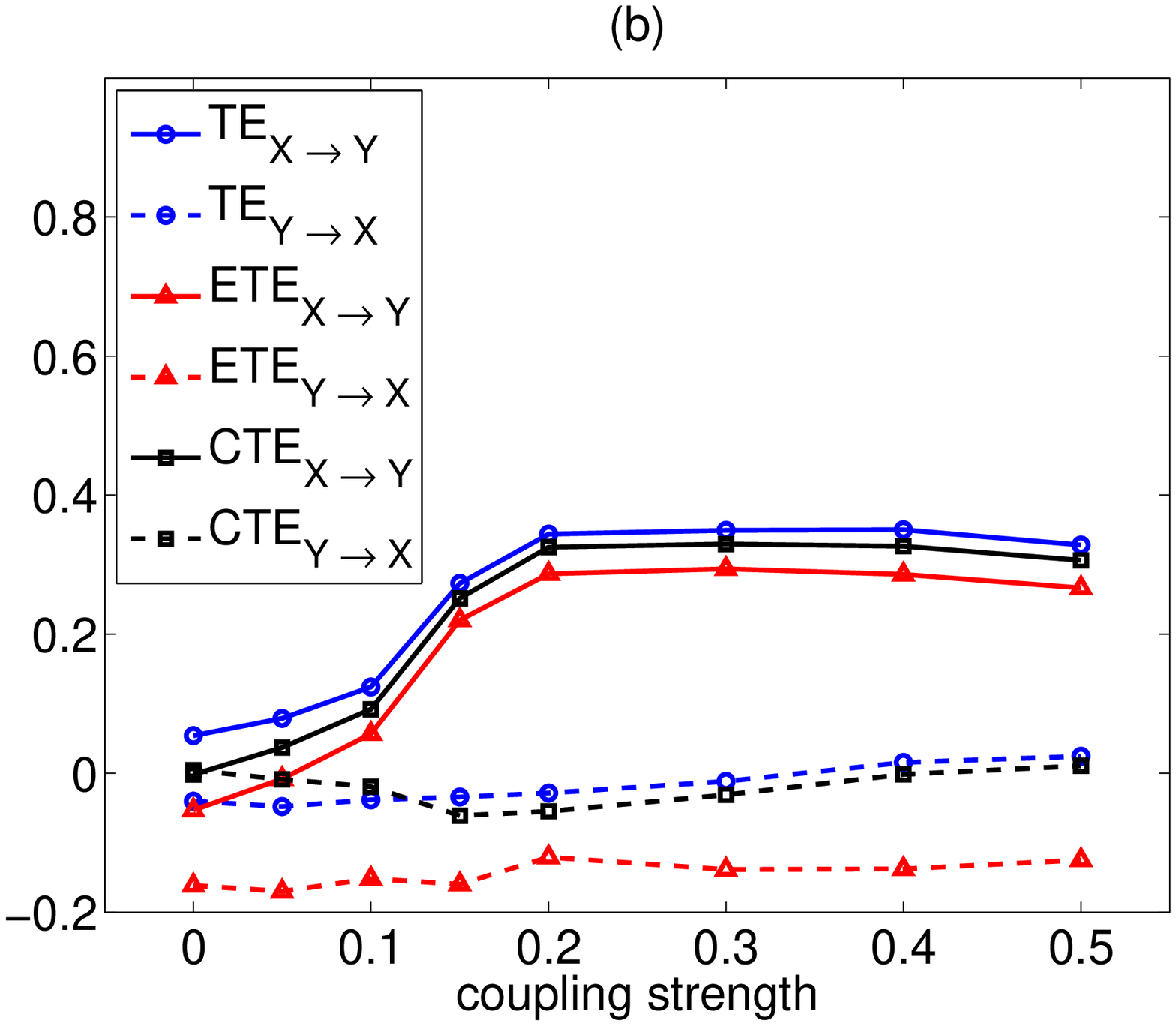}
\includegraphics[height=5.0cm,keepaspectratio]{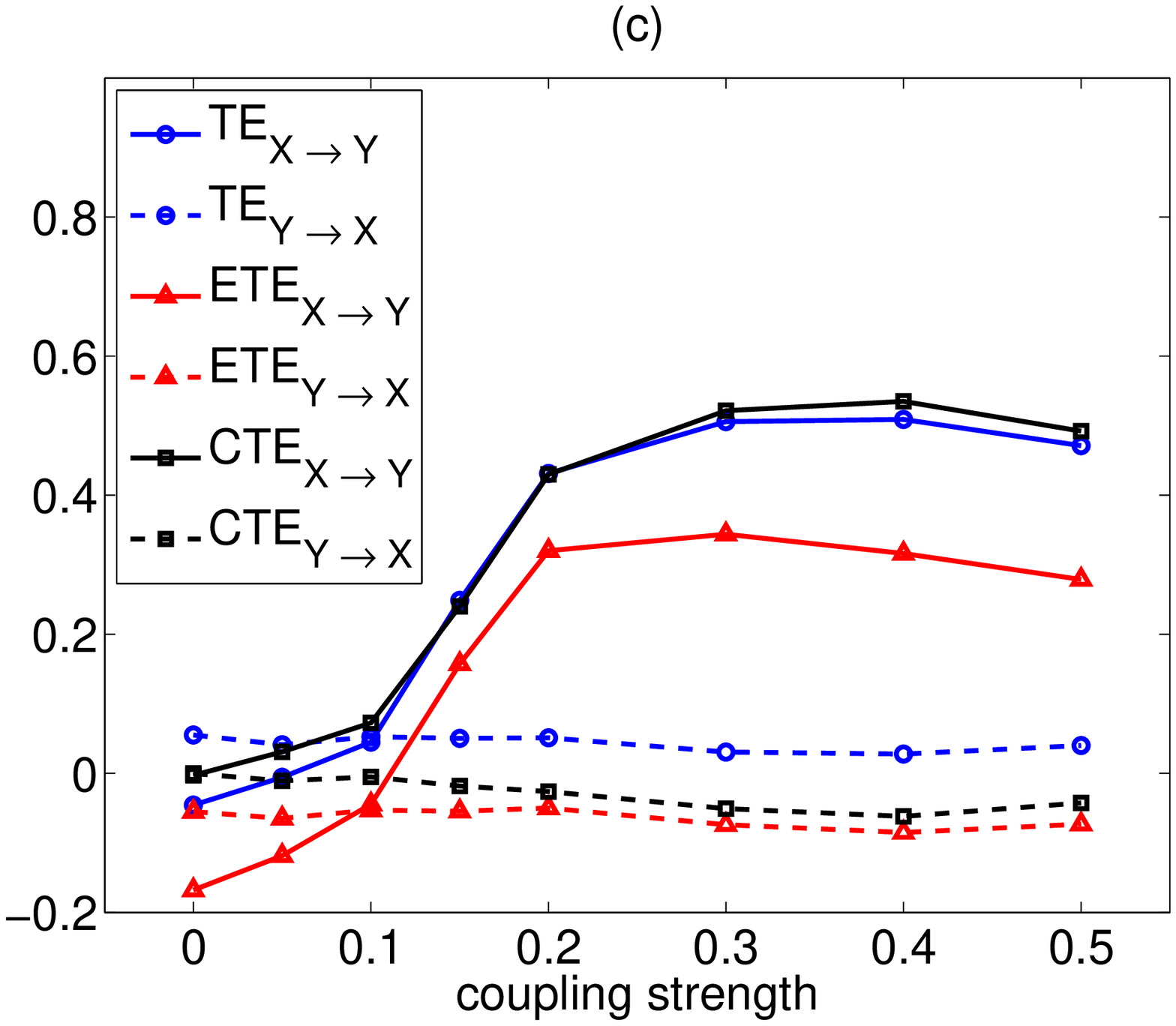}
}} \caption{(a) Mean estimated values of TE, ETE and CTE for both
directions from $100$ realizations of the noise-free
unidirectionally coupled Henon maps, with $n=512$ and $m_x=m_y=2$.
(b) and (c) as in (a) but for $m_x=2,m_y=4$ and $m_x=4,m_y=2$,
respectively.} \label{fig:TEmeasHenon}
\end{figure}

TE and mainly CTE are found to be more effective in detecting the
direction of the information flow than MCR and CMCR. The causal
effect is correctly detected in all simulations on the Henon maps.
The correct direction is preserved for all combinations of the
embedding dimensions except for $m_x=1$ and $m_y>1$, where a
spurious increase at the direction $Y \rightarrow X$ is observed.
The CTE measure is always effectively zero for both directions
when $c=0$, whereas TE tends to be positive (at cases deviating
significantly from the zero level) and ETE gives negative values
and not equal in both directions. Also, ETE is affected by the
selection of the embedding dimensions much more than TE and CTE
(see Fig.\ref{fig:TEmeasHenon}).

The three information measures turn out to be robust to noise and
the detection of the direction of interaction gets blurred only at
few combinations of embeddings dimensions for high noise levels
and short time series. This is because the variance of the
estimated measures increases with the embedding dimension and the
noise level. Thus for small coupling strengths, the distribution
of the measures in the two directions may overlap and suggest no
discrimination in the two directions.

\begin{figure} [h!]
\centerline{\hbox{
\includegraphics[height=5.0cm,keepaspectratio]{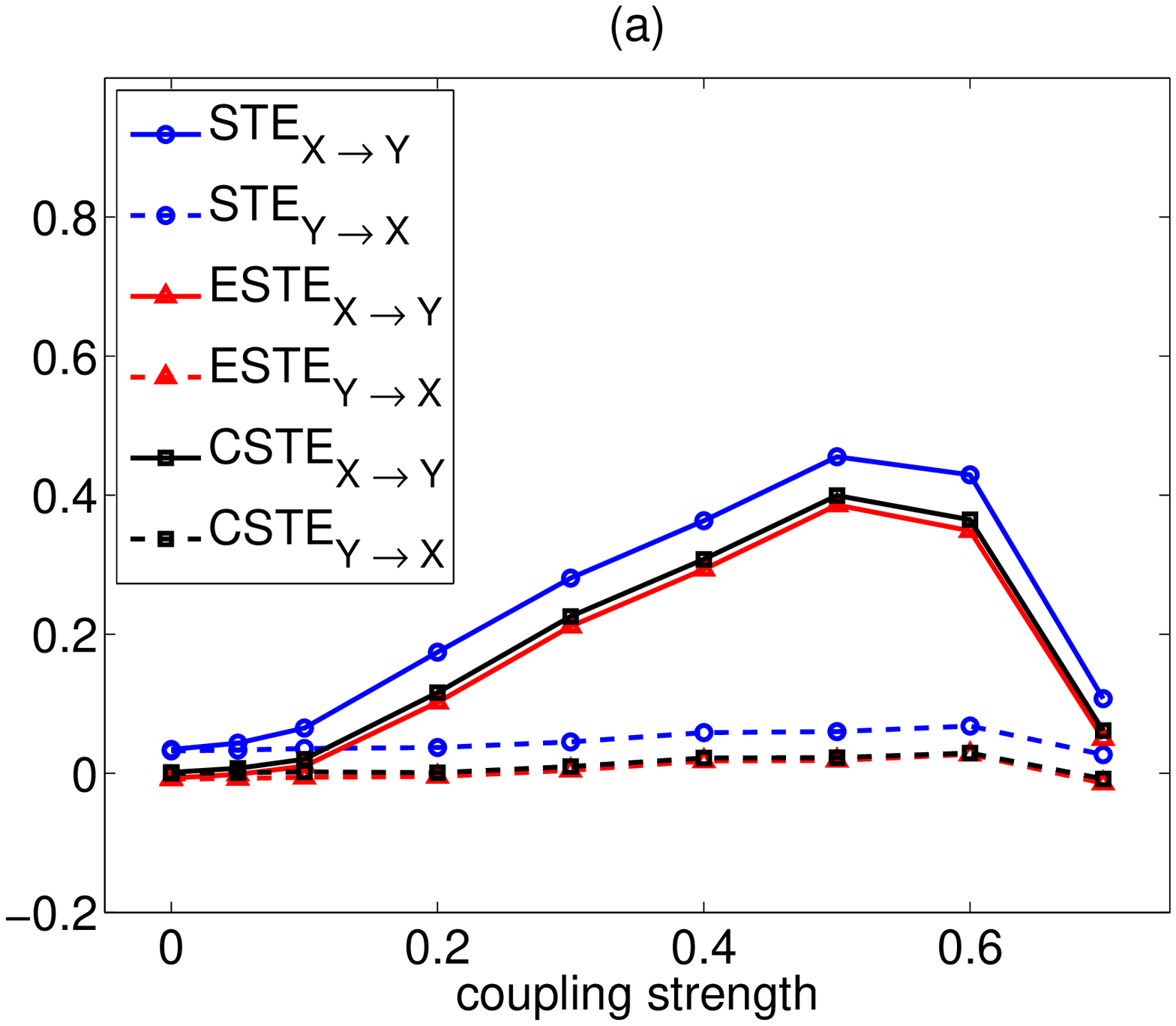}
\includegraphics[height=5.0cm,keepaspectratio]{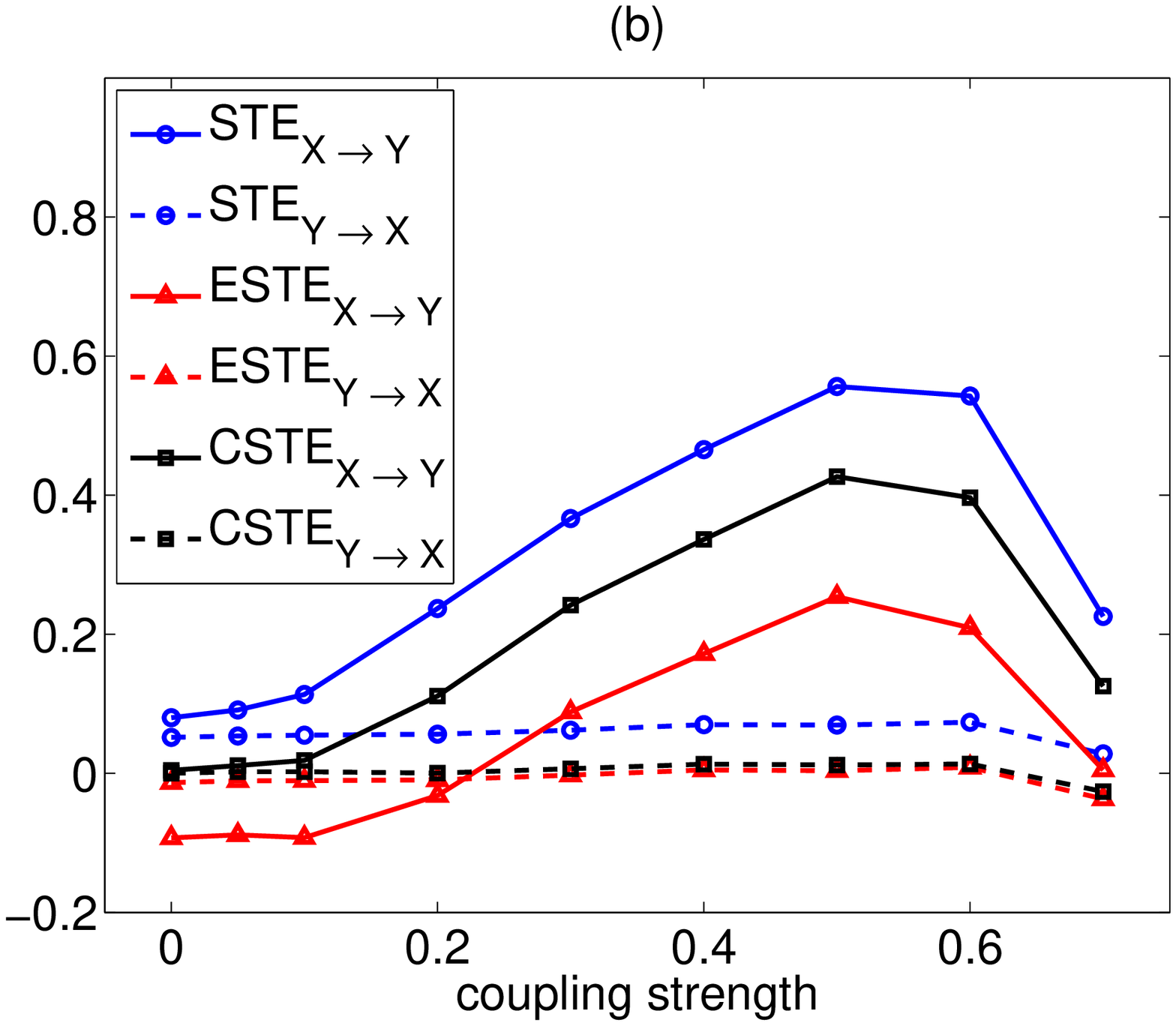}
\includegraphics[height=5.0cm,keepaspectratio]{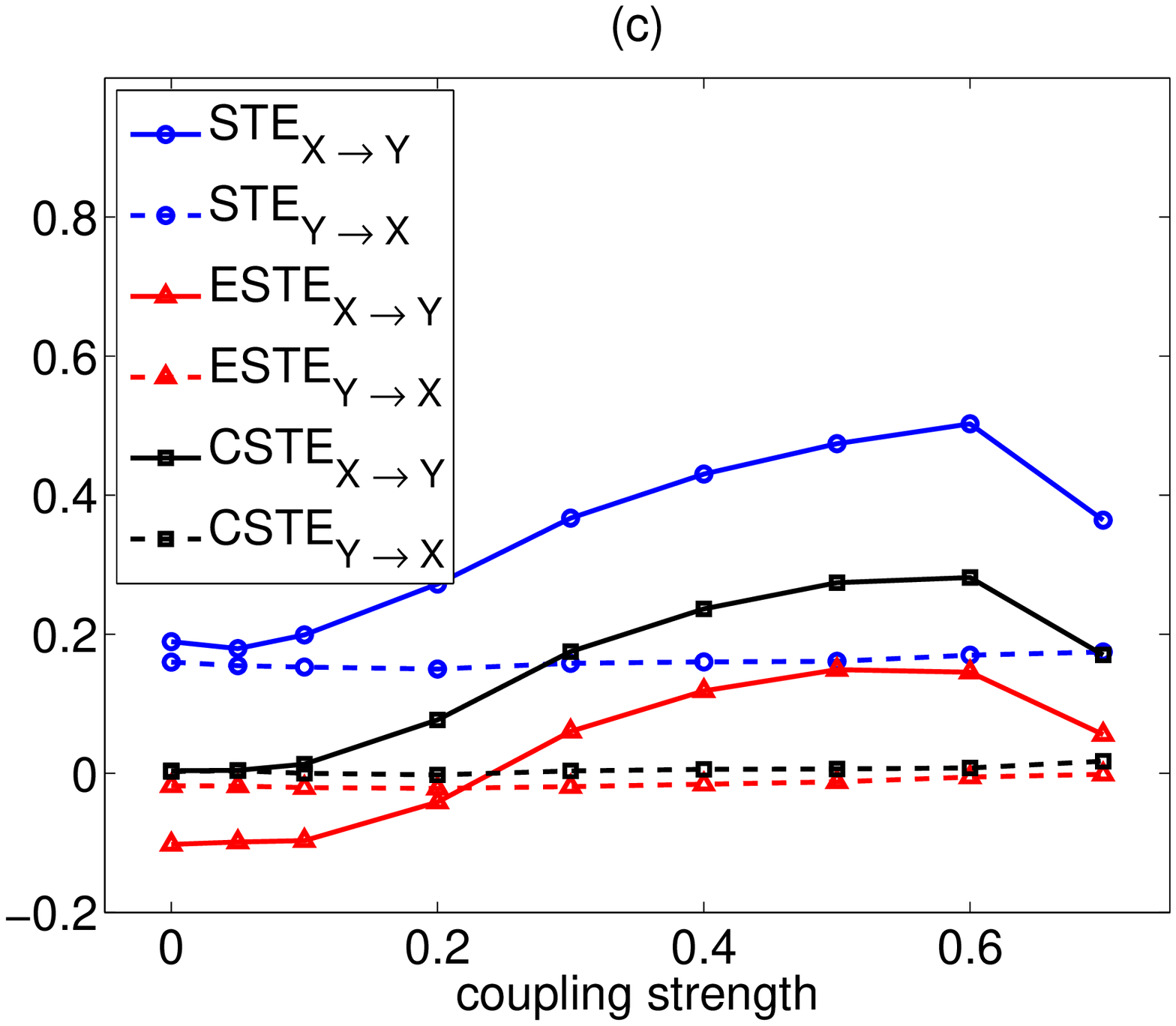}
}} \caption{(a) Mean estimated values of STE, ESTE and CSTE for
both directions from $100$ realizations of the noise-free
unidirectionally coupled Henon maps, with $n=512$ and $m_x=m_y=3$.
(b) As in (a) but for $m_x=4,m_y=3$. (c) As in (b) but for $20 \%$
noise level.} \label{fig:resHenonSTE}
\end{figure}

Symbolic transfer entropy (STE) and its corrections (CSTE and
ESTE) seem to be more affected by the selection of the embedding
dimensions than the respective TE measures. In the presence of
unidirectional coupling, STE and CSTE detect it correctly for $m_x
\geq m_y > 2$, while ESTE is significantly affected by the
embedding dimensions (see Fig.\ref{fig:resHenonSTE}a and b).
CSTE is the least sensitive to noise and gives the most consistent
results in the case of no causal effects, whereas STE is
positively biased and ESTE negatively biased (see
Fig.\ref{fig:resHenonSTE}c). The variance of the symbolic measures
is small and does not seem to increase with the addition of noise
as much as for the TE measures, so that the overlap in the two
directions for small coupling strength is much smaller. However,
for small coupling strengths, estimated values of the symbolic
measures from the two directions again overlap.

The graphical results, as those shown in
Figures~\ref{fig:henon0MCR}-\ref{fig:resHenonSTE}, are quantified
by the score of the three statistical tests. When there is no
coupling, the measures MCR and CMCR reject almost always $H_0^1$
and $H_0^2$, and reject $H_0^3$ when $m_x=m_y$, so they always
score at least 2 (see Table~\ref{tab:scoresHenon}). The reason for
the rejections is that the measures are positively biased and have
very small standard deviation, and apparently the proposed
correction of MCR can neither eliminate this bias. Though the bias
decreases with the increase of the time series length, the score
for MCR and CMCR is still at least 2 and addition of noise does
not change the score results. On the other hand, CTE scores 0 for
all combinations of embedding dimensions, even for as small time
series lengths as $n=512$, whereas TE does this only for large
$m_x, m_y$ and ETE always scores at least 2. CSTE also often
scores 0 for $m_x,m_y > 2$, while STE and ESTE perform poorly,
rejecting $H_0^1$ and $H_0^2$ mostly due to the positive and
negative bias, respectively. The proper performance of CTE and
CSTE is not substantially affected by the addition of noise.
\begin{table}
\caption{Scores for the setting of no coupling from the 100
realizations of the uncoupled Henon maps with $n=512$, for
noise-free data and $20\%$ noise (the latter values are in
parentheses).}
\begin{tabular}{|c|c|c|c|c|c|c|c|c|c|c|c|} \hline
\multicolumn{2}{|c|}{} & \multicolumn{8}{|c|}{scores, $0 \%$ noise ($20 \%$ noise)}  \\\hline %
$m_x$& $m_y$&  MCR  & CMCR &  TE   & ETE  & CTE  & STE  & ESTE & CSTE \\\hline %
  2  &  2   &  2(2) & 2(2) &  2(2) & 2(0) & 0(0) & 2(2) & 1(0) &  1(0)  \\\hline %
  2  &  3   &  3(3) & 3(3) &  3(1) & 3(0) & 0(0) & 2(3) & 0(0) &  1(0)  \\\hline %
  2  &  4   &  3(3) & 3(3) &  3(2) & 3(2) & 0(0) & 3(2) & 2(3) &  1(1)  \\\hline %
  2  &  5   &  3(3) & 3(3) &  3(2) & 3(2) & 0(0) & 2(3) & 2(2) &  0(0)  \\\hline %
  3  &  2   &  3(3) & 3(3) &  3(0) & 3(0) & 0(1) & 2(2) & 2(0) &  3(0)  \\\hline %
  3  &  3   &  2(2) & 2(2) &  2(0) & 2(0) & 0(0) & 2(2) & 2(2) &  0(0)  \\\hline %
  3  &  4   &  3(3) & 3(3) &  2(0) & 3(3) & 0(0) & 2(3) & 2(3) &  0(0)  \\\hline %
  3  &  5   &  3(3) & 3(3) &  3(0) & 3(3) & 0(0) & 3(3) & 3(3) &  0(0)  \\\hline %
  4  &  2   &  3(3) & 3(3) &  3(0) & 3(2) & 0(1) & 3(2) & 3(2) &  0(1)  \\\hline %
  4  &  3   &  3(3) & 3(3) &  2(2) & 3(3) & 0(1) & 3(3) & 2(3) &  2(0)  \\\hline %
  4  &  4   &  2(2) & 2(2) &  0(0) & 2(2) & 0(0) & 2(2) & 2(2) &  0(0)  \\\hline %
  4  &  5   &  3(3) & 3(3) &  0(2) & 3(2) & 0(2) & 3(2) & 3(3) &  0(0)  \\\hline %
  5  &  2   &  3(3) & 3(3) &  3(2) & 3(3) & 0(1) & 2(3) & 2(2) &  1(0)  \\\hline %
  5  &  3   &  3(3) & 3(3) &  3(0) & 3(3) & 0(0) & 3(3) & 3(3) &  2(0)  \\\hline %
  5  &  4   &  3(3) & 3(3) &  0(0) & 3(2) & 0(2) & 3(2) & 3(3) &  0(0)  \\\hline %
  5  &  5   &  3(2) & 3(2) &  0(2) & 2(0) & 0(2) & 2(2) & 2(3) &  0(2)  \\\hline %
\end{tabular}
\label{tab:scoresHenon}
\end{table}

For the second setting of the presence of unidirectional coupling,
the measures perform rather similarly. CMCR and MCR give scores 2
or 3, meaning that besides $H_0^1$ also $H_0^2$ is rejected,
erroneously due to positive bias, and at cases $H_0^3$ is rejected
as well. $H_0^2$ is often rejected by the information measures due
to either positive bias (TE) or negative bias (ETE and sometimes
CTE). ETE gives systematically negative values, while CTE might
have a slightly negative mean at some cases, however its estimated
values are around zero. CSTE turns out to outperform all the other
measures and gives a proper score 2 (only $H_0^2$ is not rejected)
and it is also robust against noise, even for high level of noise
($20 \%$).

\subsection{Results on the unidirectionally coupled Mackey-Glass systems}
\begin{figure} [h!]
\centerline{\hbox{
\includegraphics[height=5.0cm,keepaspectratio]{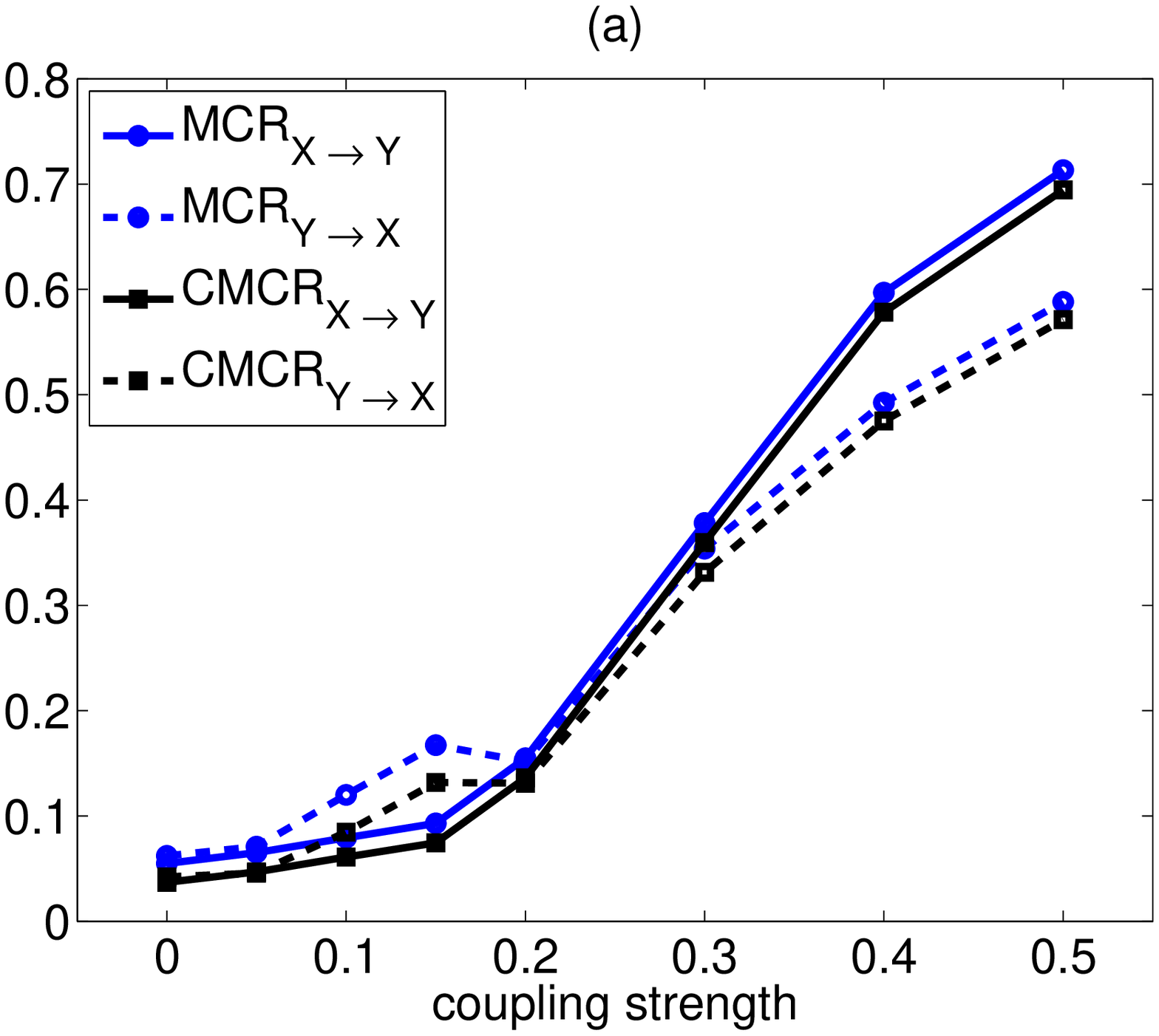}
\includegraphics[height=5.0cm,keepaspectratio]{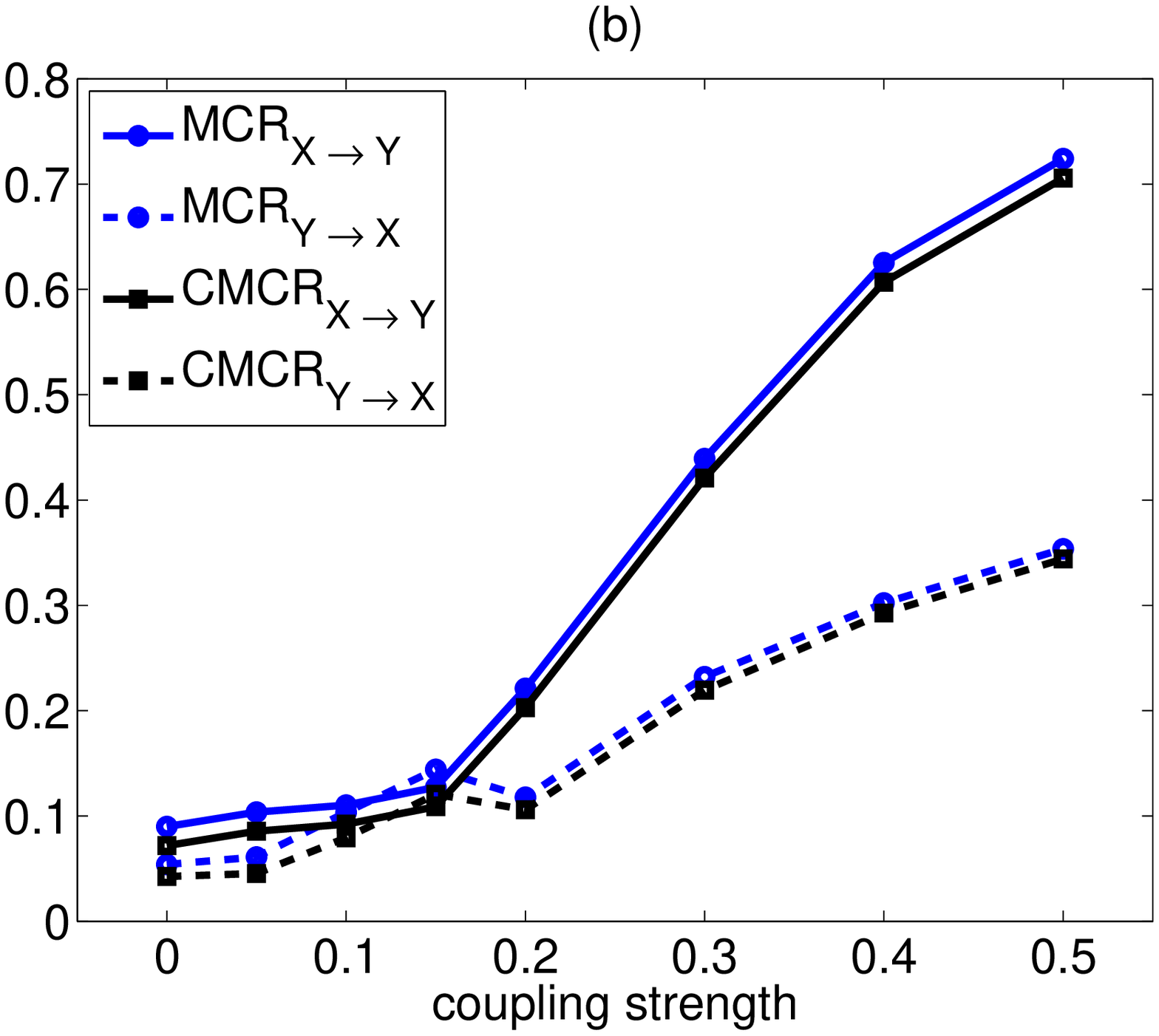}
}} \caption{(a) Mean estimated values of MCR and CMCR for both
directions from $100$ realizations of the noise-free
unidirectionally coupled Mackey-Glass systems with $\Delta_x=30$
and $\Delta_y=100$, $n=2048$ and $m_x=m_y=3$. (b) As in (a) but
for $m_x=4, m_y=3$.} \label{fig:MGlassMCR}
\end{figure}

For the unidirectionally coupled Mackey-Glass systems, the MCR
measures increase also in the opposite direction with the coupling
strength for all embedding dimensions. The MCR and CMCR values are
larger in the correct direction only for $m_x \leq m_y$, whereas
for $m_x=m_y$ they increase close together in both directions (see
Fig.\ref{fig:MGlassMCR}). Addition of noise worsens the
performance of the MCR measures. It is noteable that for
$\Delta_x=\Delta_y$, MCR and CMCR point to the wrong direction of
interaction.

\begin{figure} [h!]
\centerline{\hbox{
\includegraphics[height=5.0cm,keepaspectratio]{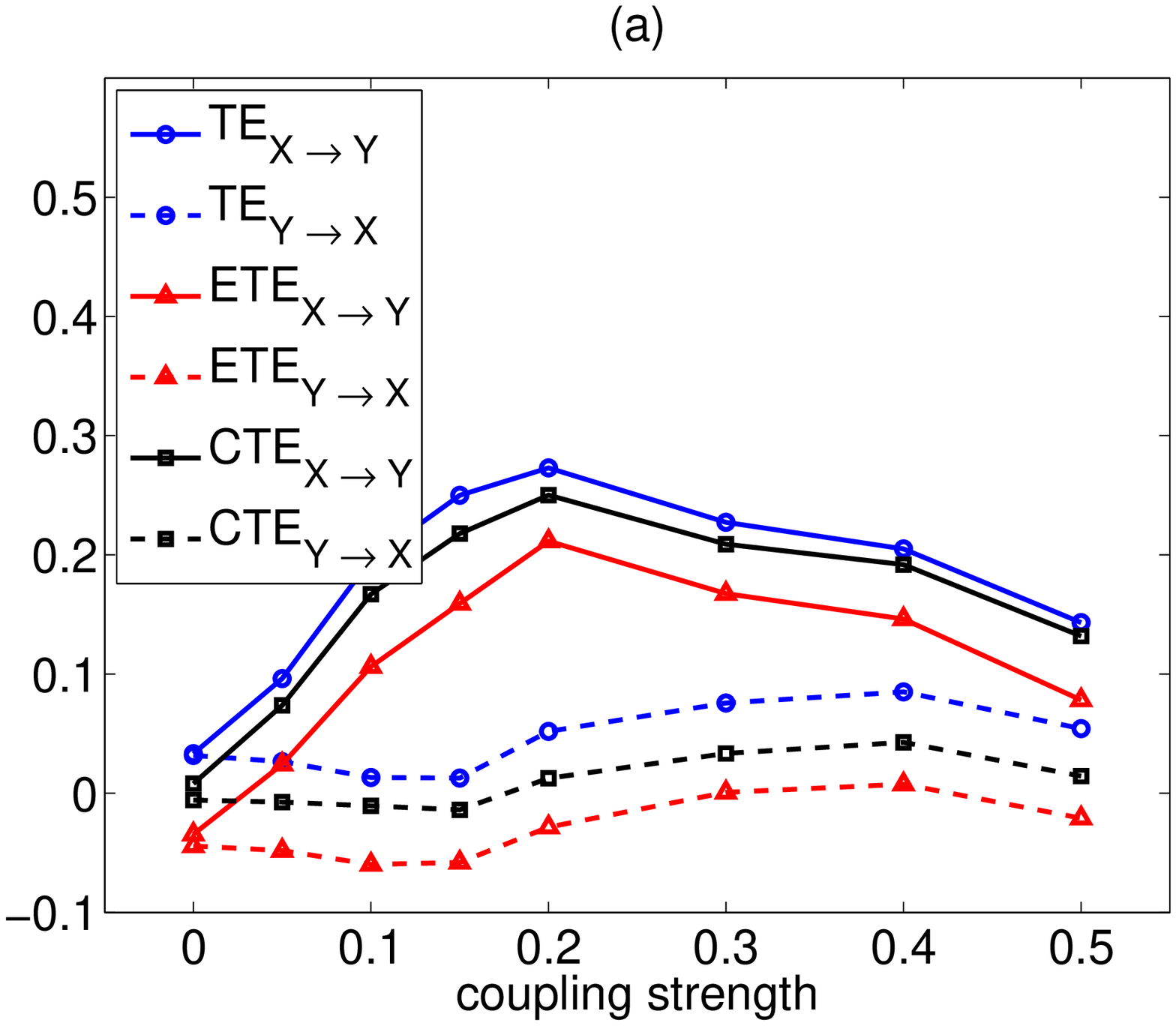}
\includegraphics[height=5.0cm,keepaspectratio]{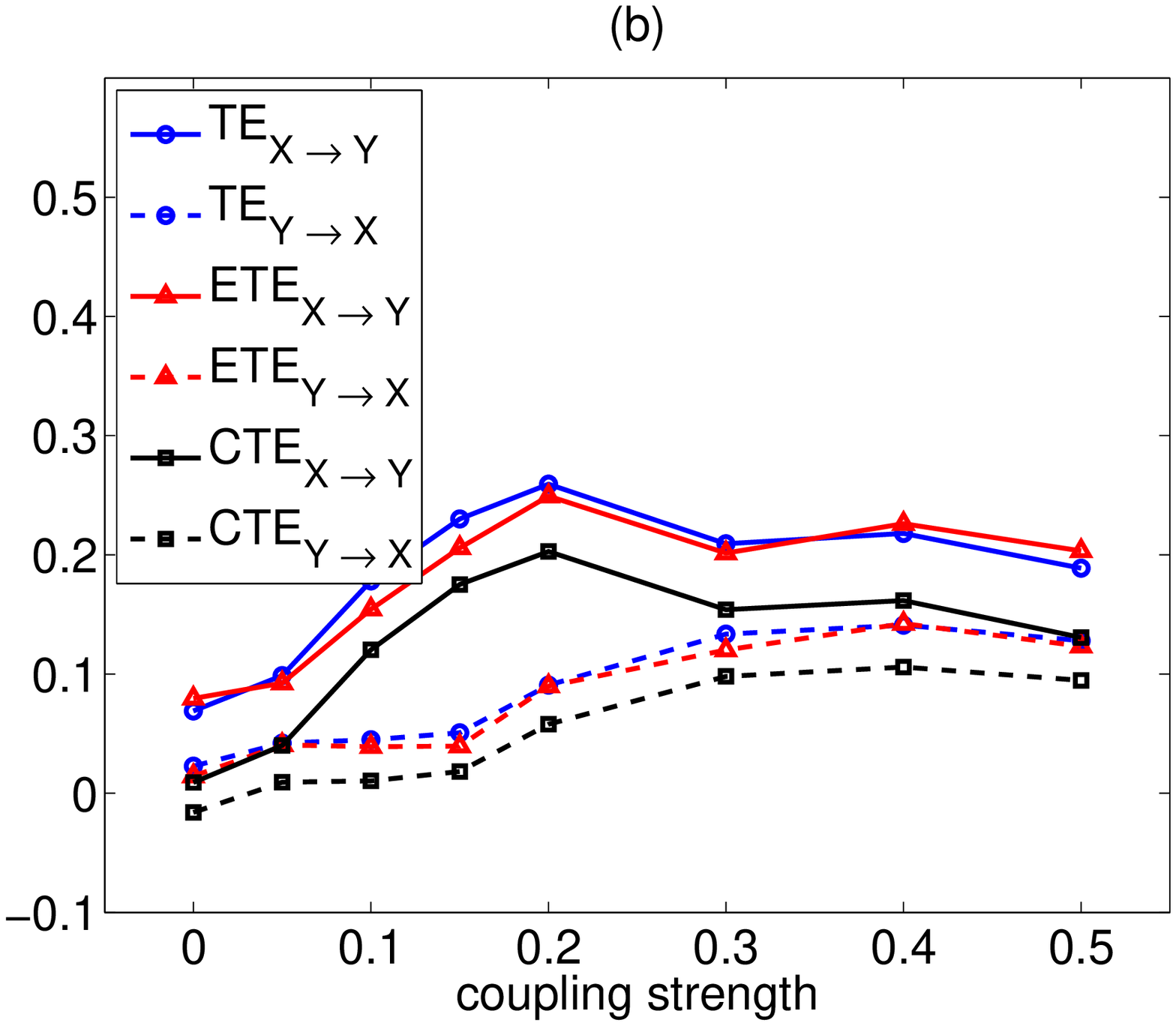}
}} \centerline{\hbox{
\includegraphics[height=5.0cm,keepaspectratio]{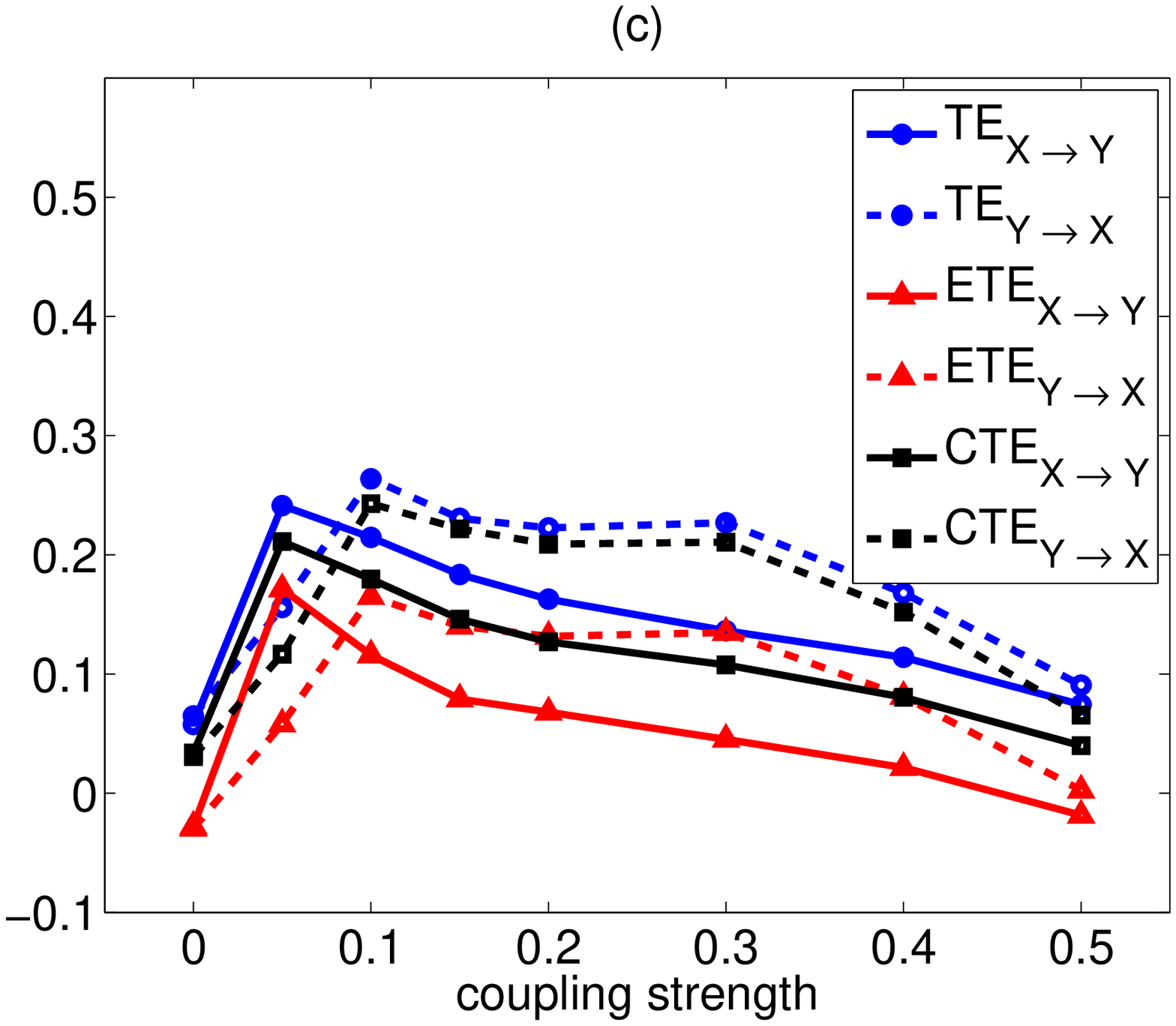}
\includegraphics[height=5.0cm,keepaspectratio]{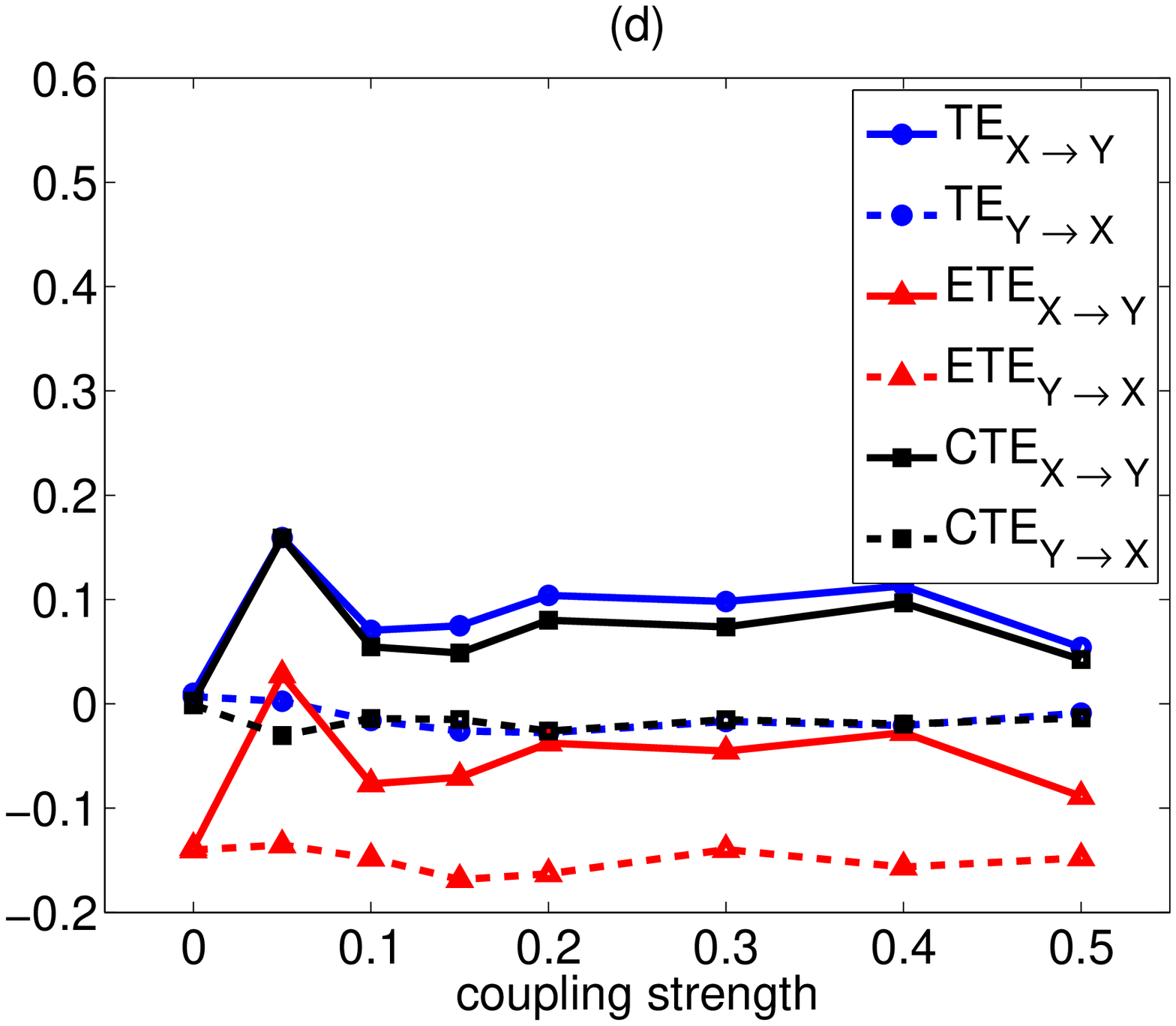}
}} \caption{(a) Mean estimated values of TE, ETE and CTE for both
directions from $100$ realizations of the noise-free
unidirectionally coupled Mackey-Glass systems ($\Delta_x=30$,
$\Delta_y=100$) with $n=2048$ and $m_x=4$ and $m_y=3$. (b) As in
(a) but for $20 \%$ noise level. (c) and (d) As in (a) but for
$\Delta_x=\Delta_y=17$, $m_x=m_y=3$ and $m_x=m_y=5$,
respectively.} \label{fig:MGlassTEmeas}
\end{figure}

The transfer entropy measures are also significantly affected by
the embedding dimension. The direction of the causal effect is
generally detected with all measures when $m_x \geq m_y$, but
fails when $m_x$ is much smaller than $m_y$. For example, for
$\Delta_x=30$ and $\Delta_y=100$, all measures detect the correct
driving effect when $m_x \geq m_y$, as shown in
Fig.\ref{fig:MGlassTEmeas}a for $m_x=4$, $m_y=3$, contrary to the
MCR measures shown for the same setup in Fig.\ref{fig:MGlassMCR}b.
For systems with $\Delta_x=\Delta_y$, the detection is problematic
even for $m_x=m_y$, and the stronger the coupling the larger
$m_x=m_y$ is needed to be detected. This feature is shown in
Fig.\ref{fig:MGlassTEmeas}c and d for $\Delta_x=\Delta_y=17$,
where $\mbox{TE}_{X\rightarrow Y} > \mbox{TE}_{Y\rightarrow X}$
holds only for very weak coupling when $m_x=m_y=3$ (see
Fig.\ref{fig:MGlassTEmeas}c), and in order to achieve
$\mbox{TE}_{X\rightarrow Y}>\mbox{TE}_{Y\rightarrow X}$ also for
stronger coupling $m_x=m_y$ has to be increased to 5 (see
Fig.\ref{fig:MGlassTEmeas}d). In all cases, TE, CTE and ETE show
the same signature (almost parallel lines in
Fig.\ref{fig:MGlassTEmeas}), but CTE attains best the zero level
at $c=0$, whereas TE is slightly positively biased for certain
embedding dimensions and ETE is negatively biased. Addition of
noise does not change these structures but decreases their mean
value and increase their variance, particularly for large
embedding dimensions (see Fig.\ref{fig:MGlassTEmeas}b).

\begin{figure} [h!]
\centerline{\hbox{
\includegraphics[height=5.0cm,keepaspectratio]{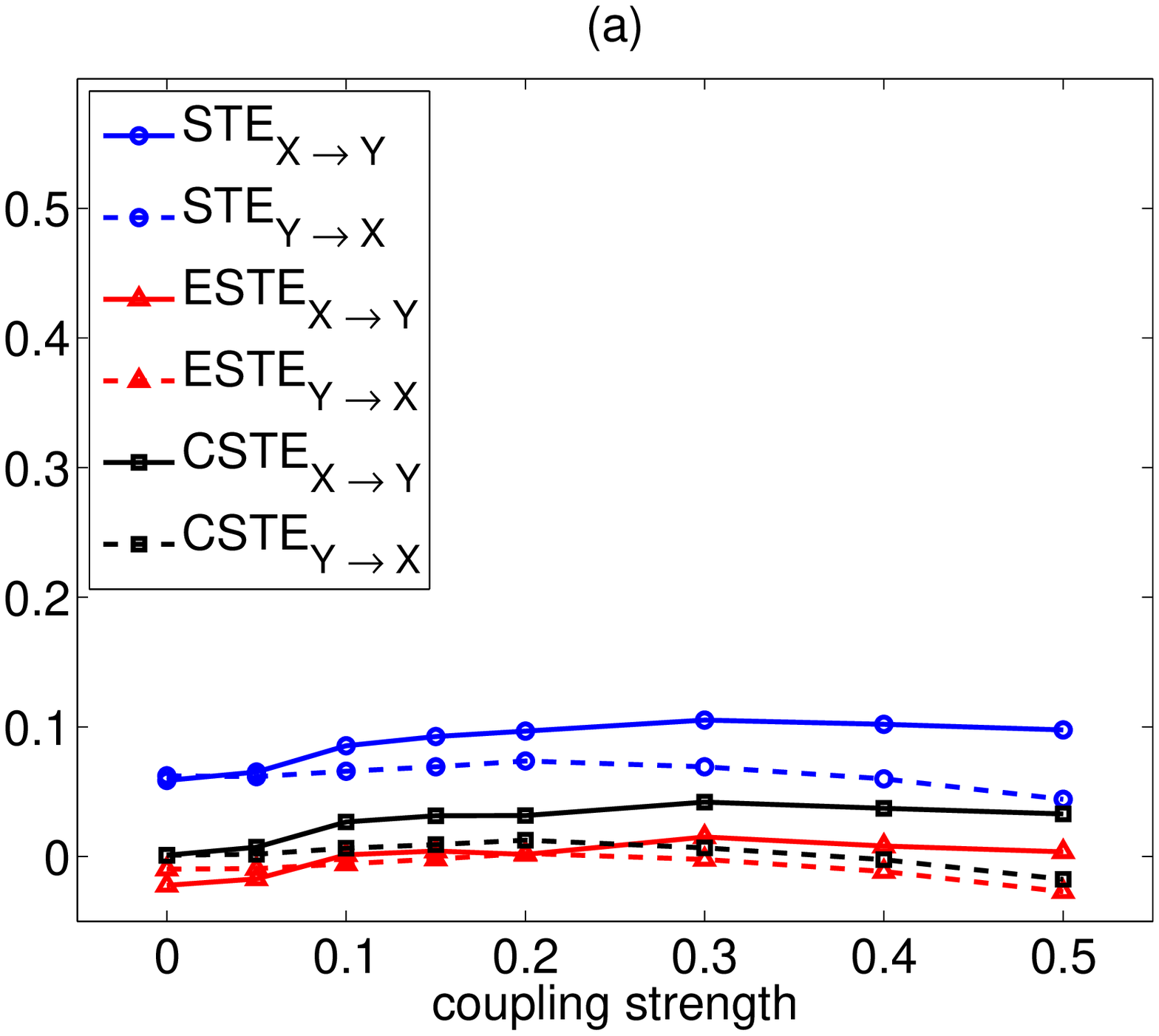}
\includegraphics[height=5.0cm,keepaspectratio]{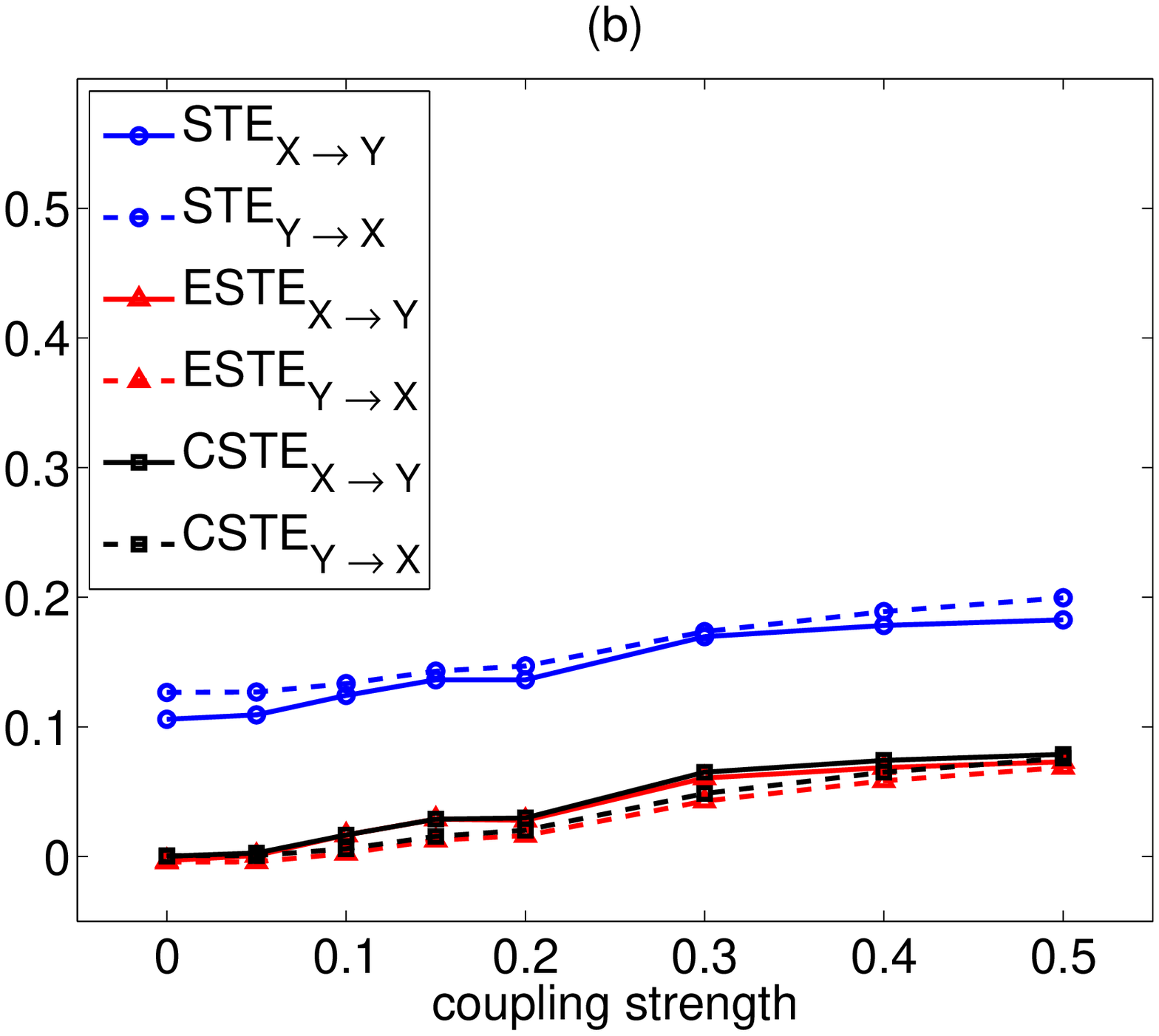}
\includegraphics[height=5.0cm,keepaspectratio]{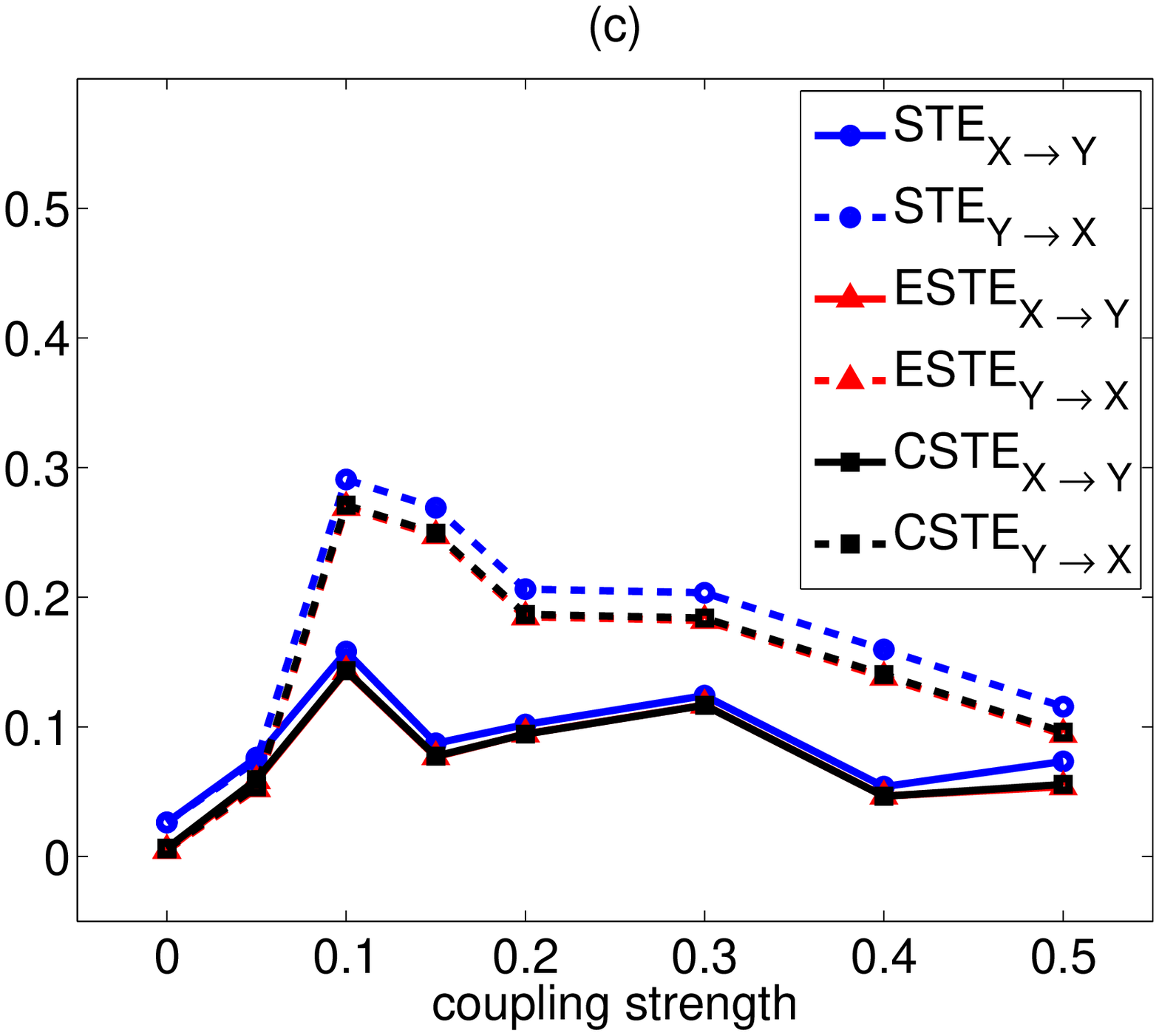}
}} \caption{The panels are as for the three first panels of
Fig.\ref{fig:MGlassTEmeas} but for the symbolic transfer entropy
measures.} \label{fig:MGlassSTEmeas}
\end{figure}

The symbolic transfer entropy measures depend on the embedding
dimensions more than the transfer entropy measures, and fail more
often to detect the correct causal effect, as can be seen from the
comparison of the results on TE measures in
Fig.\ref{fig:MGlassTEmeas}a, b and c and STE measures in
Fig.\ref{fig:MGlassSTEmeas} for the same simulation setup.
However, CSTE gives values around zero for any selection of
embedding dimensions in case of no causal effects.

Regarding the formal hypothesis tests, for the uncoupled
Mackey-Glass systems, MCR and CMCR scored high in the first
setting of no coupling and rejected almost always $H_0^1$ and
$H_0^2$ for the Mackey-Glass system (see
Table~\ref{tab:MGd17r30scores} for $\Delta_x=17$, $\Delta_y=30$
and $m_x=m_y$).
\begin{table}[h!]
\caption{Scores for the setting of no coupling of the information
measures, from the 100 realizations of the uncoupled Mackey-Glass
system ($\Delta_x=17$, $\Delta_y=30$) with $n=2048$, for the
noise-free case and for $20 \%$ noise level.}
\begin{tabular}{|c|c|c|c|c|c|c|c|c|} \hline
\multicolumn{1}{|c|}{} & \multicolumn{8}{|c|}{scores, $0 \%$ noise ($20 \%$ noise)} \\\hline %
$m_x=m_y$ & MCR  & CMCR &  TE  & ETE  & CTE  & STE  & ESTE & CSTE   \\\hline %
 2        & 3(3) & 3(3) & 3(2) & 3(0) & 3(0) & 2(2) & 2(0) & 2(0)   \\\hline %
 3        & 3(3) & 3(3) & 2(2) & 3(0) & 2(0) & 2(2) & 2(0) & 2(0)   \\\hline %
 4        & 3(3) & 3(3) & 2(0) & 3(2) & 0(0) & 3(3) & 3(3) & 1(1)   \\\hline %
 5        & 3(3) & 3(3) & 2(0) & 3(3) & 0(0) & 3(3) & 3(3) & 0(0)   \\\hline %
 6        & 3(3) & 3(3) & 2(1) & 3(2) & 1(1) & 3(3) & 3(3) & 1(0)   \\\hline %
 7        & 3(3) & 3(3) & 1(2) & 3(2) & 1(2) & 3(2) & 3(3) & 1(0)   \\\hline %
\end{tabular}
\label{tab:MGd17r30scores}
\end{table}
CTE and CSTE scored overall worse than for the Henon map, but
still better than their respective counterpart, TE, ETE and STE,
ESTE, respectively. For example, for $\Delta_x=30$ and
$\Delta_y=100$, CSTE scored 0 in most of the combinations of $m_x$
and $m_y$, while the other information measures performed poorly
scoring mostly 3 or 2.

\begin{figure} [h!]
\centerline{\hbox{
\includegraphics[height=5.0cm,keepaspectratio]{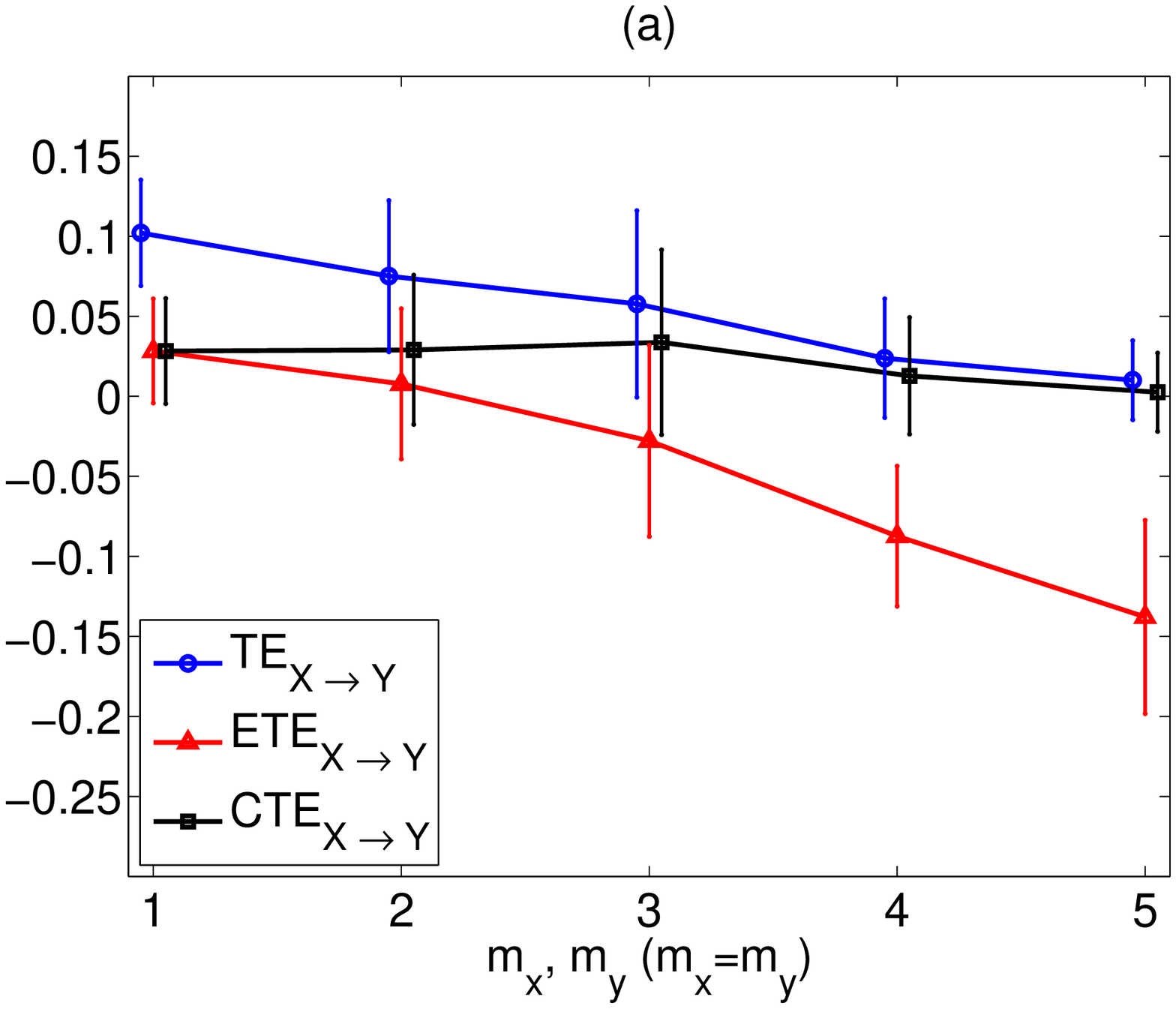}
\includegraphics[height=5.0cm,keepaspectratio]{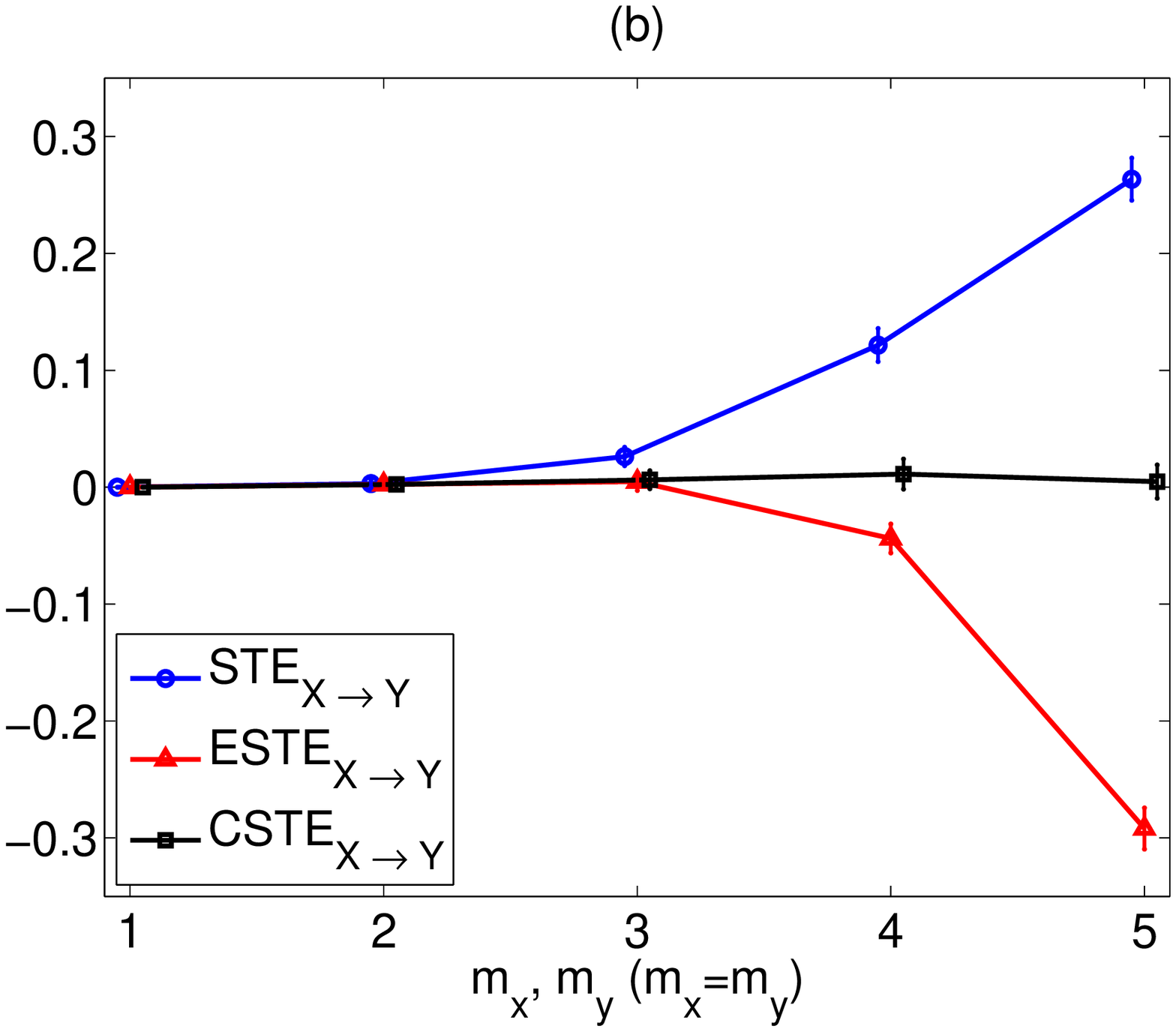}
\includegraphics[height=5.0cm,keepaspectratio]{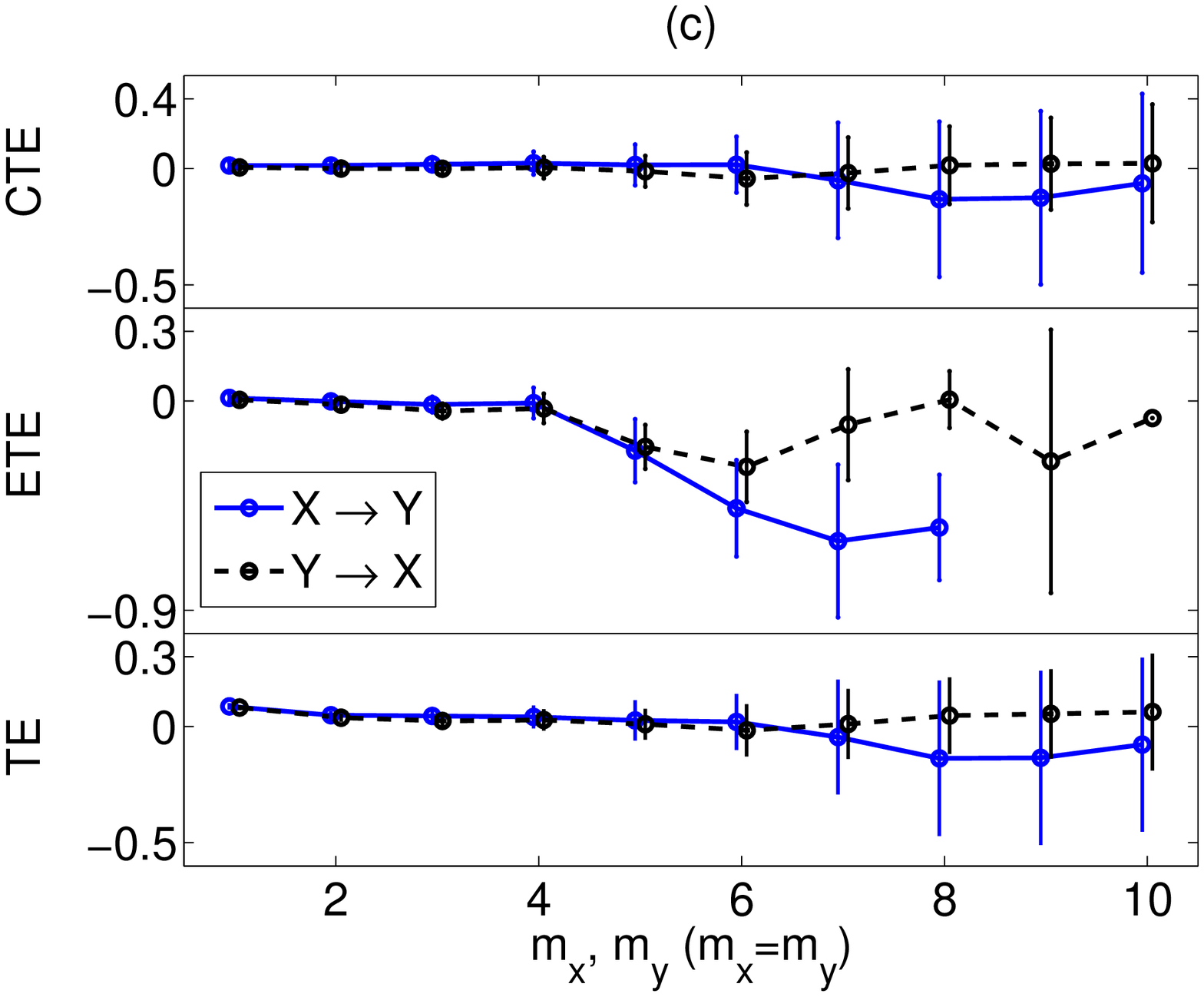}
}} \caption{(a) Mean TE, ETE and CTE (error bars denote the
standard deviation) from $100$ realizations of the noise-free
uncoupled Mackey-Glass systems with $\Delta_1=\Delta_2=17$,
$n=2048$, and for varying $m_x=m_y$. Only the one direction is
shown as the two systems are the same. (b) As in (a) but for STE,
ESTE and CTE. (c) As in (a) but for $\Delta_1=30, \Delta_2=100$
and the two directions are shown, as given in the legend, with a
measure at each panel. The drawn points and error bars for each
measure are slightly displaced along the x-axis to facilitate
visualization.}
 \label{fig:MGlassc0}
\end{figure}

For all scenarios of complexity of the coupled Mackey-Glass
systems, STE obtains significant positive values also for $c=0$
when $m_x < m_y$, while ESTE obtains often negative values. CSTE
follows the same dependence on $c$ as STE but displaced so that
CSTE falls at the zero level for $c=0$. It is interesting that in
the presence of noise, STE gets even larger values for $c=0$ and
increases faster for $c > 0$, and CSTE has the same course with
$c$ as STE but starts at the zero level for $c=0$. We illustrate
this nice property of CTE and CSTE for $c=0$ as a function of
$m_x$, where $m_x = m_y$ in Fig.\ref{fig:MGlassc0}.
First, we note that STE measures have much less variance than the
respective TE measures and attain the zero level for small $m_x =
m_y$ (in Fig.\ref{fig:MGlassc0}a only the distribution of CTE
contains zero for varying $m_x = m_y$). In
Fig.\ref{fig:MGlassc0}c, where the systems have different
complexity, ETE gets more affected by the individual system
complexity as the embedding dimension increases, TE and CTE do not
differ much in the two directions, but only CTE is at the zero
level.

In the presence of unidirectional coupling, MCR and CMCR score at
least 2 as the first and second H$_0$ are rejected, meaning that
the measures are positively biased and the estimated values of the
measures increase at the same level in both directions. The latter
does not occur with the information measures and $H_0^3$ is almost
always rejected, as well as $H_0^1$ (the correct direction of
coupling). In this task, the corrected measures (CTE and CSTE)
performed similarly to the original measures.

The simulations on the coupled Mackey-Glass systems showed that
CTE and CSTE improve the performance of the original measures,
giving values closer to zero when the systems are uncoupled.
Similarly to the coupled Henon maps, the optimal selection for
embedding dimensions is $m_x=m_y$. For $m_x > m_y$, the dynamics
of the driving system are over-represented giving larger TE and
STE values in the correct direction, whereas for $m_x < m_y$ the
opposite effect is observed decreasing TE and STE for $X
\rightarrow Y$ and increasing TE and STE for $Y \rightarrow X$, so
that for very small $m_x$ the measure values are even larger for
the wrong direction $Y \rightarrow X$. Though CTE and CSTE
decrease the positive bias due to uneven representation of the
systems when $m_x \neq m_y$ they cannot remove it completely.
Another bias that cannot be vanished by the correction of the
transfer entropy measures is due to the individual dynamics, which
persist for $m_x=m_y$. The bias turns out to be larger when the
two systems have identical individual dynamics, i.e.
$\Delta_x=\Delta_y$. We found that for all three
$\Delta_x=\Delta_y$ values we tested for, only small coupling
strength $c$ could be detected correctly using small $m_x=m_y$ and
for larger $c$ the embedding dimension should be increased with
the complication that it may be too large for the given time
series length. We attribute this to the similarity of the
trajectories of the driving and response system, even when they
are not in phase, so that a larger time window length from each
trajectory is required to detect differences (in terms of
entropies) that reveal the driving effect. When $\Delta_x \neq
\Delta_y$, as driving increases the shape of the trajectories of
the response system gets closer to that of the driving system and
therefore the driving effect can be detected better even for small
$m_x=m_y$. Indeed this was the case for all combinations of
$\Delta=100$ and any of $\Delta=17, 30$ (at any order of driving
and response). For the pair $(\Delta_x=17, \Delta_y=30)$ the
difference in the individual dynamics was smaller and the
detection of the correct driving effect for $c>0.2$ required that
$m_x=m_y$ be as large as 6, while for the pair $(\Delta_x=30,
\Delta_y=17)$ even larger $m_x=m_y$ was required.

\subsection{Results on the coupled nonlinear stochastic
system}
For the coupled nonlinear stochastic system of three variables the
driving effects $X \rightarrow Y$, $Y \rightarrow Z$ and $X
\rightarrow Z$ take place at lag one, so we expect that $m_x = m_y
=1$ be sufficient for all pairs of variables. However, MCR gives
larger values at the correct driving effect $X\rightarrow Y$
(meaning any of $X \rightarrow Y$, $Y \rightarrow Z$ and $X
\rightarrow Z$) for larger embedding dimensions $m_x \geq m_y$ and
with a large bias that decreases with the increase of time series
length. CMCR reduces the MCR values but does not attain the zero
level when evaluated for no coupling or opposite driving effect.

\begin{figure} [h!]
\centerline{\hbox{
\includegraphics[height=5.0cm,keepaspectratio]{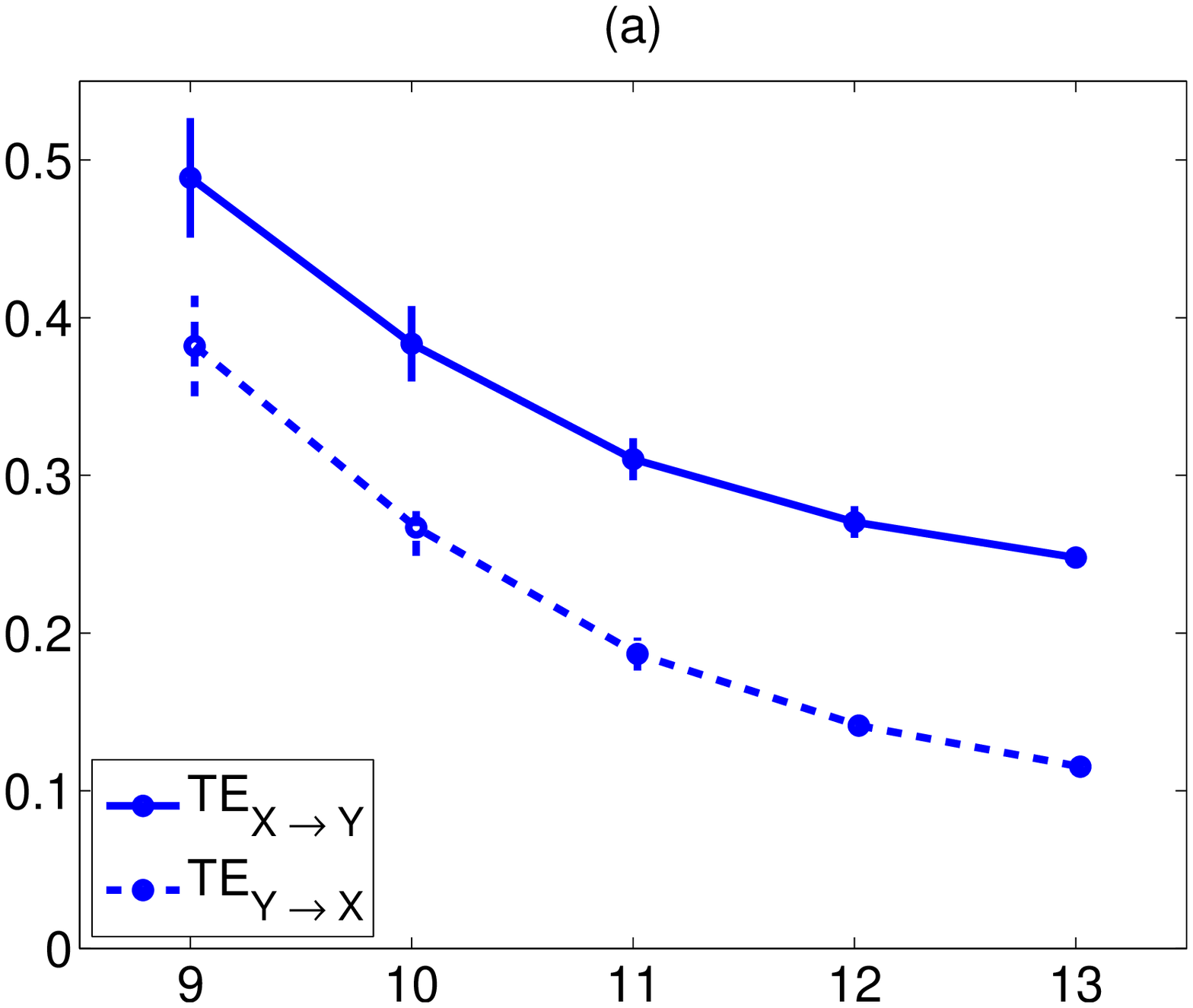}
\includegraphics[height=5.0cm,keepaspectratio]{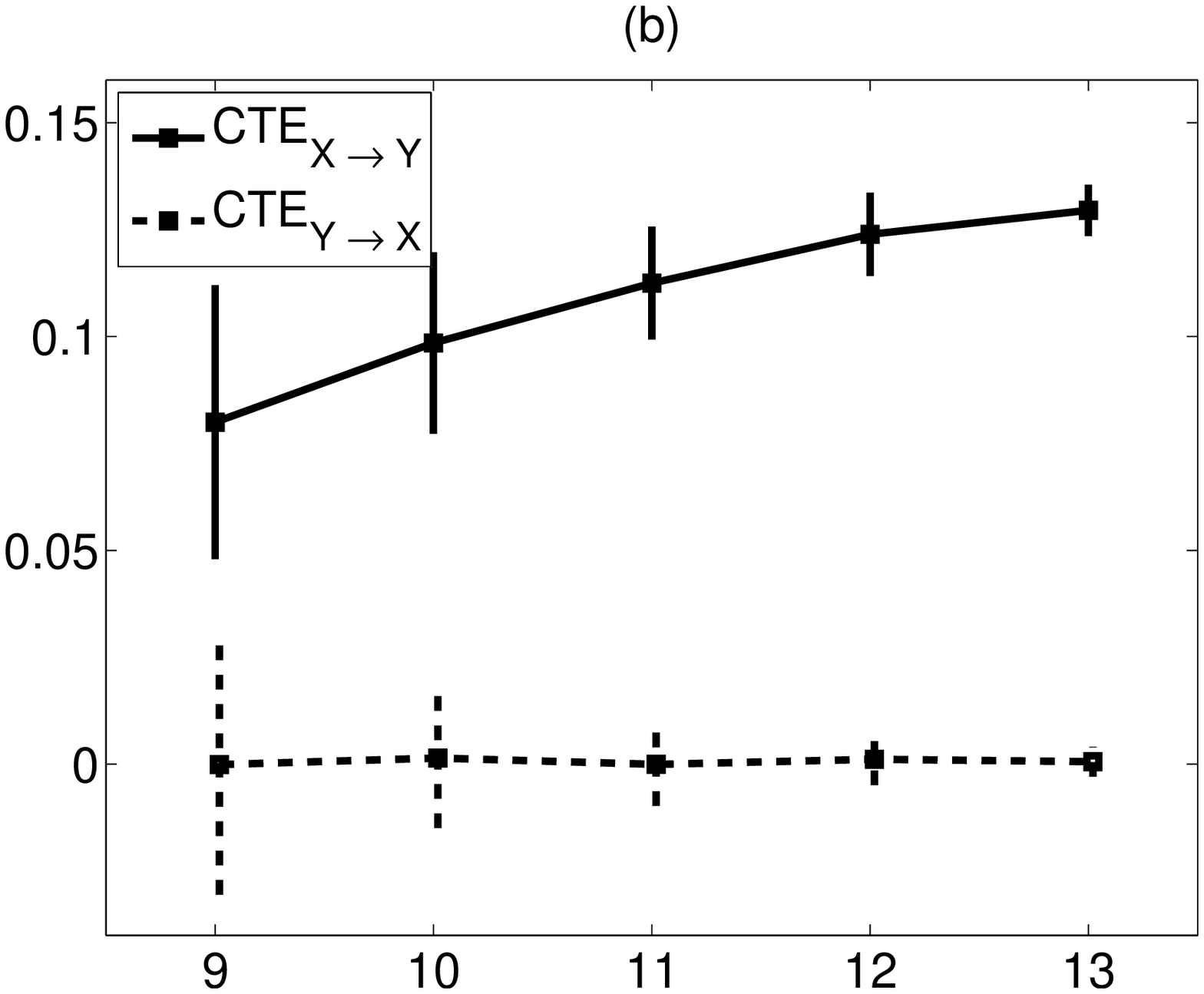}
}} \centerline{\hbox{
\includegraphics[height=5.0cm,keepaspectratio]{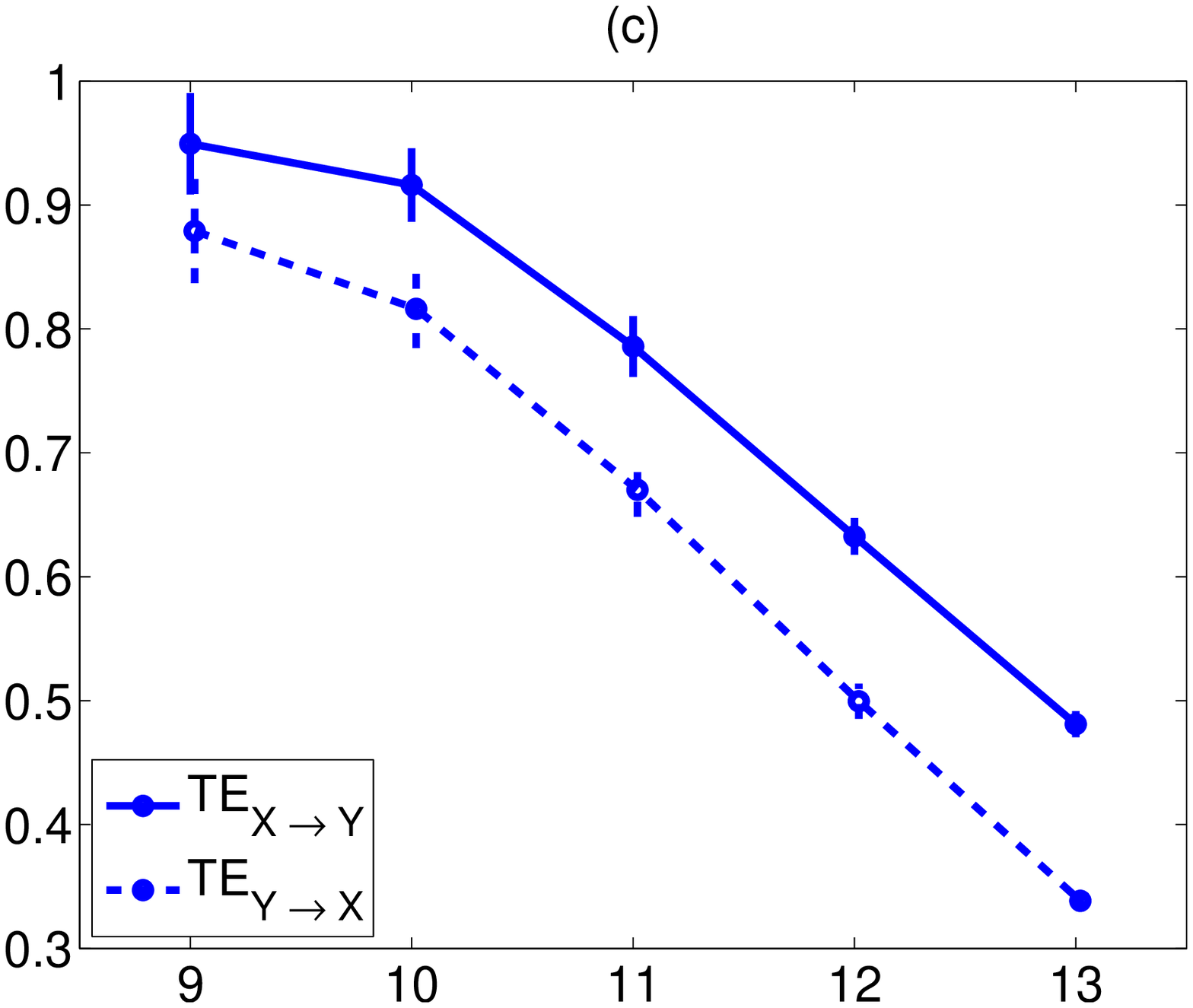}
\includegraphics[height=5.0cm,keepaspectratio]{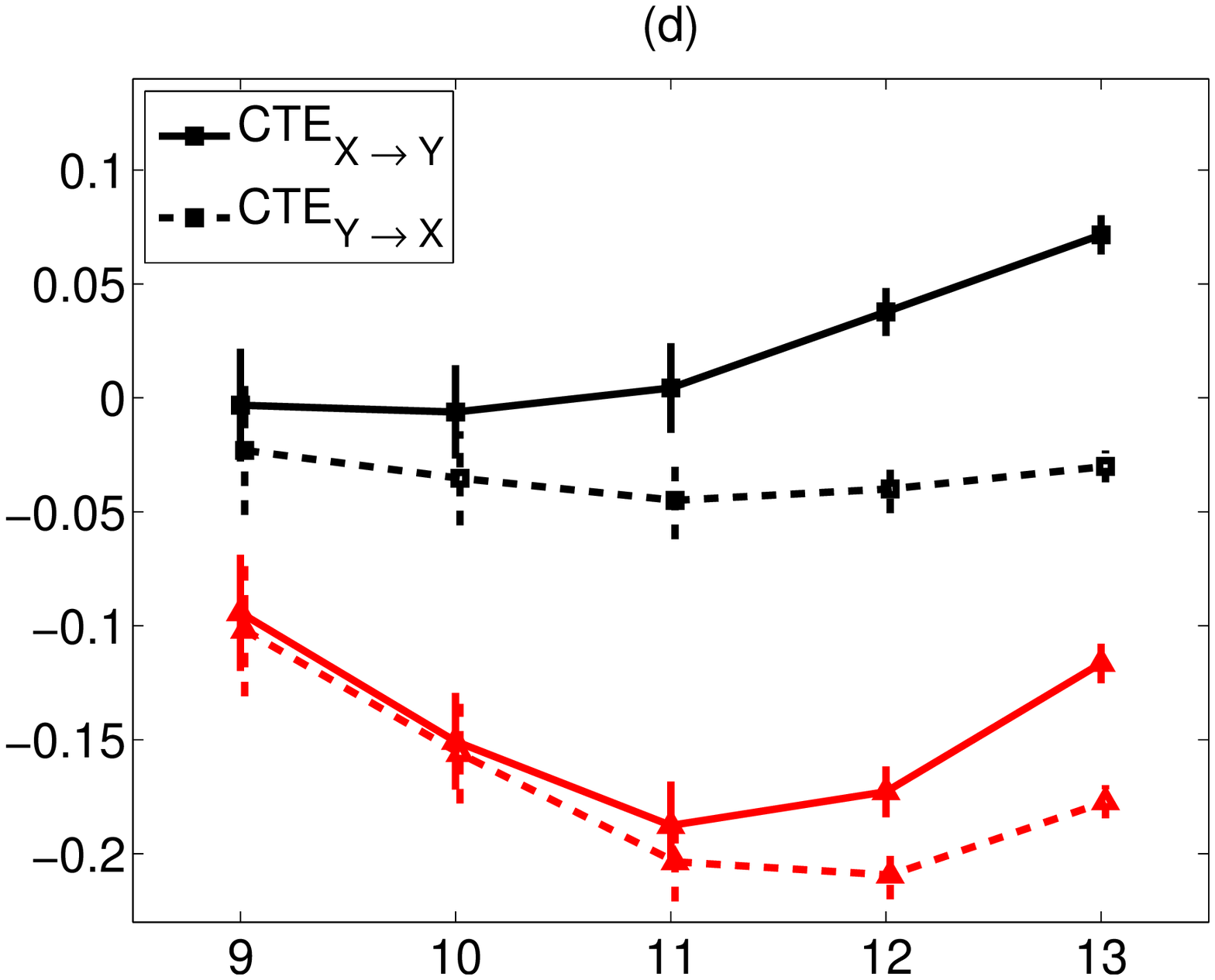}
}}
 \caption{(a) Mean TE (error bars denote the
standard deviation) from $100$ realizations of the coupled
stochastic system vs time series length $\log_2 n$, for the
variables $X$ and $Y$, and $m_x=m_y=1$. (b) As in (a) but for CTE.
(c) As in (a) but for $m_x=m_y=2$. (d) As in (b) but for CTE and
ETE and $m_x=m_y=2$. The drawn points and error bars for each
measure are slightly displaced along the x-axis to facilitate
visualization.} \label{fig:stochastic}
\end{figure}

TE is also positively biased for all embedding dimensions and one
can only observe the correct driving from the relative difference
in the two directions. Though TE decreases with the increase of
time series length, it stays positive also for the opposite
driving effect (for the pair $(X,Y)$ of the stochastic systems see
Fig.\ref{fig:stochastic}a for $m_x=m_y=1$ and
Fig.\ref{fig:stochastic}c for $m_x=m_y=2$).
CTE resolves this problem reducing the bias in both directions so
that CTE for the opposite driving effect is at the zero level
(Fig.\ref{fig:stochastic}b for $m_x=m_y=1$ and
Fig.\ref{fig:stochastic}d for $m_x=m_y=2$). Note that for
$m_x=m_y=1$ CTE reduces to ETE, whereas for $m_x=m_y=2$ ETE is
different and goes negative obtaining smaller relative difference
between the two directions as for the other systems. The same
results are observed for the other two variable pairs.

The situation with STE and its correction is similar to TE, but
starting at $m_x=m_y=2$, with the only difference that for the
pair $(X,Z)$ the correct driving $X \rightarrow Z$ is less evident
as for this case it is nonlinear and weaker.

\subsection{Comparison to other types of surrogates}
\begin{figure} [h!]
\centerline{\hbox{
\includegraphics[height=5.0cm,keepaspectratio]{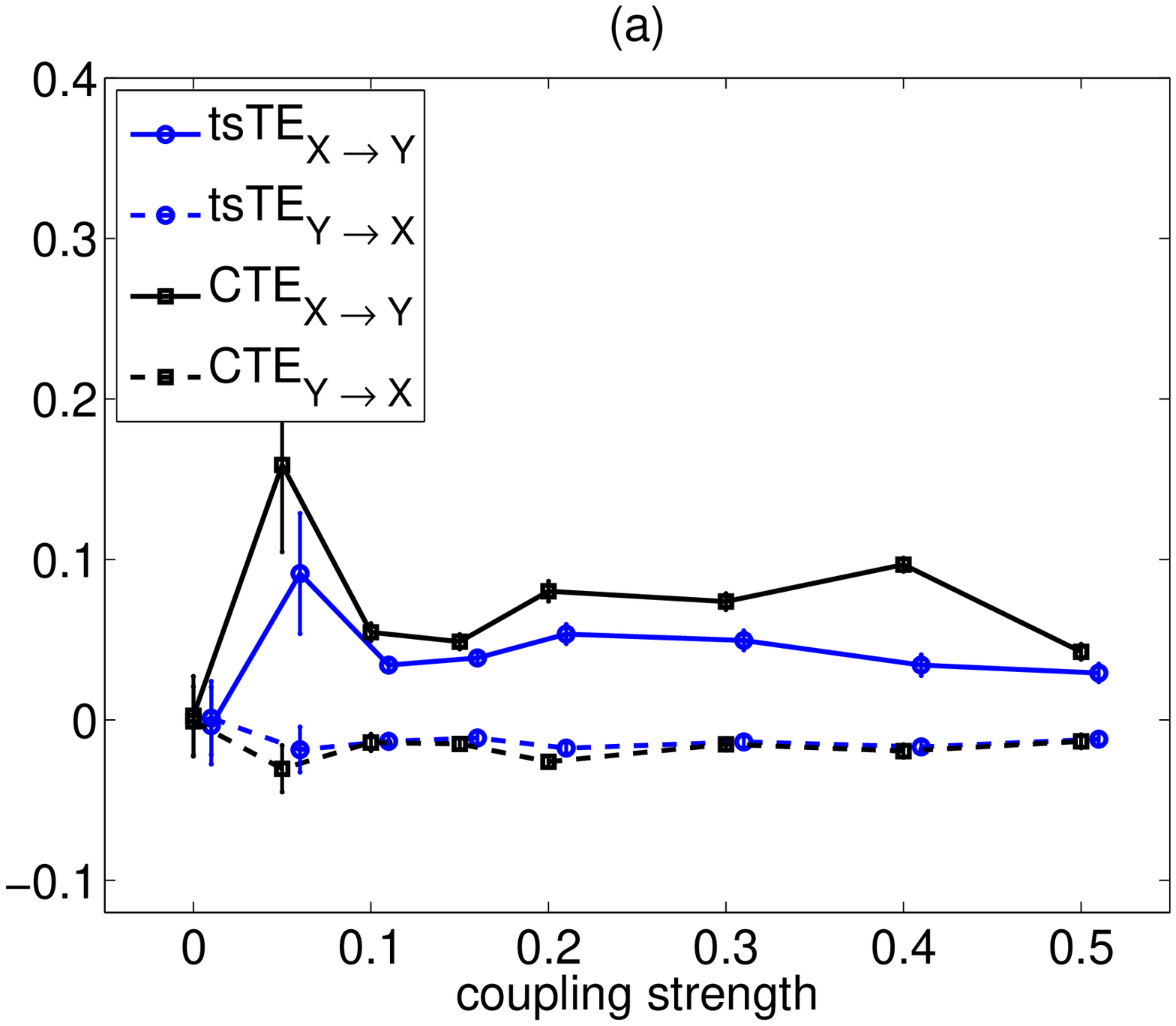}
\includegraphics[height=5.0cm,keepaspectratio]{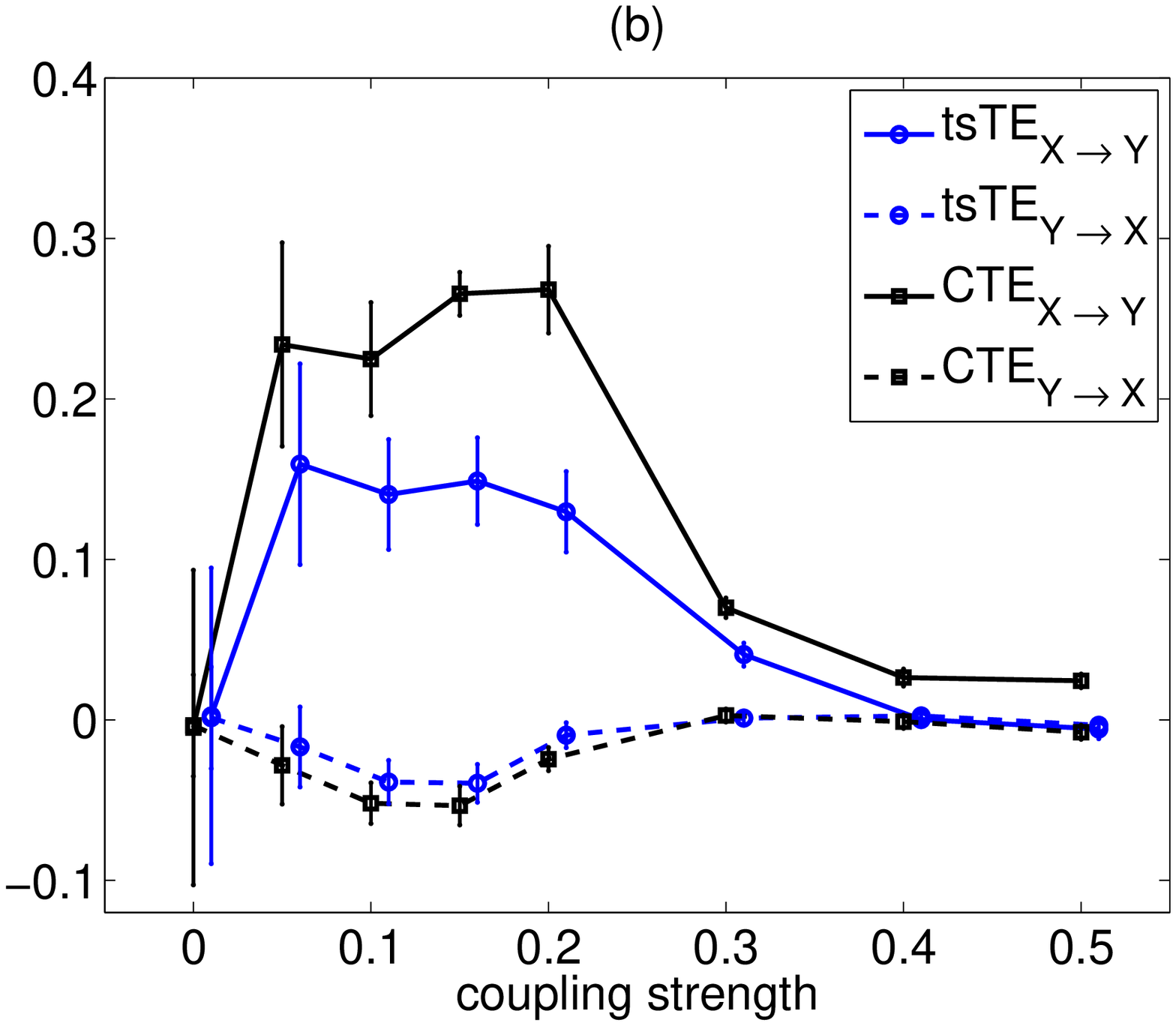}
\includegraphics[height=5.0cm,keepaspectratio]{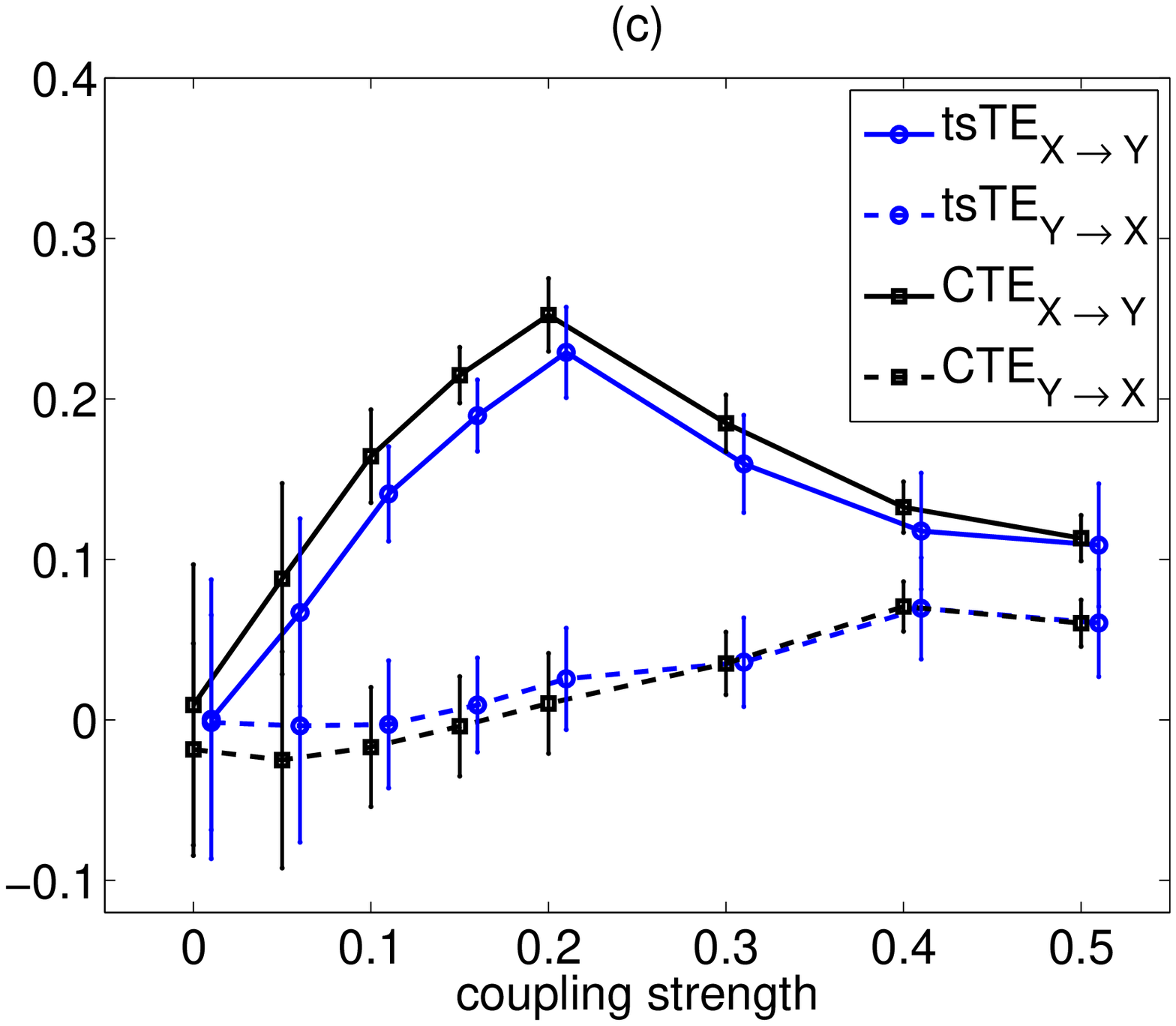}
}} \caption{(a) Mean CTE and TE from the time-shifted surrogates,
denoted as tsTE (error bars denote the standard deviation) from
$100$ realizations of the noise-free Mackey-Glass systems with
$\Delta_1=\Delta_2=17$, $n=2048$, and for $m_x=m_y=5$. (b) and (c)
are as (a) but for $\Delta_1=17, \Delta_2=30$ and for
$\Delta_1=30, \Delta_2=100$, respectively. The curves and error
bars for each measure are slightly displaced along the x-axis to
facilitate visualization.}
 \label{fig:tsCTE}
\end{figure}

We compare the corrected measures defined in terms of random
shuffling of the reconstructed points to other surrogates data
schemes, i.e. twin surrogates \cite{Thiel06} and time-shifted
surrogates \cite{Faes08}. We concentrate on the TE measure, but
our simulations with STE produced similar results. Measures using
the twin or time-shifted surrogates are estimated as the
difference of TE on the original bivariate time series and the
mean TE on $M$ surrogates. CTE turns out to perform the same as
for the two types of surrogates or even better at cases. Although
all three measures (using shuffled reconstructed vectors, twin
surrogates and time-shifted surrogates) establish the zero level
for the direction $Y \rightarrow X$, CTE is larger for $X
\rightarrow Y$ for the whole range of $c>0$. Representative
examples are given in Fig.\ref{fig:tsCTE} for the time-shifted
surrogates and the Mackey-Glass system.
We note here that twin surrogates have the highest computational
cost because of the long computation time in constructing the
surrogates.

\section{Application to EEG}
\label{sec:Application}
The measures considered in the simulation study are evaluated on
two scalp preictal EEG records of 25 channels (system 10--20 with
added low rows) and one intracranial EEG preictal record of 28
channels in a grid. We want to evaluate how the measures detect
changes in the interactions of any pair of channels from the early
to the late preictal state. The first extracranial EEG record is
for a generalized tonic clonic seizure and the other for a left
back temporal lobe epilepsy. No specific artifact removal method
was applied but to attain better source derivation at small
cortical regions, for each EEG channel, the mean EEG of the four
neighboring channels was subtracted \cite{Hjort75}. The pairs of
transformed EEG that were used for the estimation of the measures
are: central left (C3) vs right (C4), temporal left (T7) vs right
(T8), frontal left (F3) vs right (F4) and parietal left (P3) vs
right (P4). For the intracranial EEG, the pairs of channels were
either from the same brain area (two left frontal (LTP-1 vs
LTP-3), two left temporal (LST-1 vs LST-3), two left occipital
channels (ROT-1 vs ROT-3) and the same for the right side (RTP-1
vs RTP-3, RST-1 vs RST-3, ROT-1 vs ROT-3), or from opposite brain
areas (a left and right frontal (LTP-1 vs RTP-1), temporal (LST-1
vs RST-1) and occipital channels (LOT-1 vs ROT-1).

The data windows are from 4 hours to 3 hours before the seizure
onset (early preictal state) and the last one hour before the
seizure onset (late preictal state). Each one hour long data
window is split to 120 successive non-overlapping segments of 30
sec and the causality measures are estimated for the channel pairs
at each segment and for both directions. As the sampling frequency
is 100 Hz, each 30 sec segment consists of 3000 data points. For
the estimation of the measures, the embedding dimensions are
$m_x=m_y=3$ and $m_x=m_y=5$, the time horizon is $h=5$, the lags
are $\tau_x=\tau_y=5$ and the radius is $r=0.2$ times the standard
deviation of the data.

\begin{figure} [h!]
\centerline{\hbox{
\includegraphics[height=5.0cm,keepaspectratio]{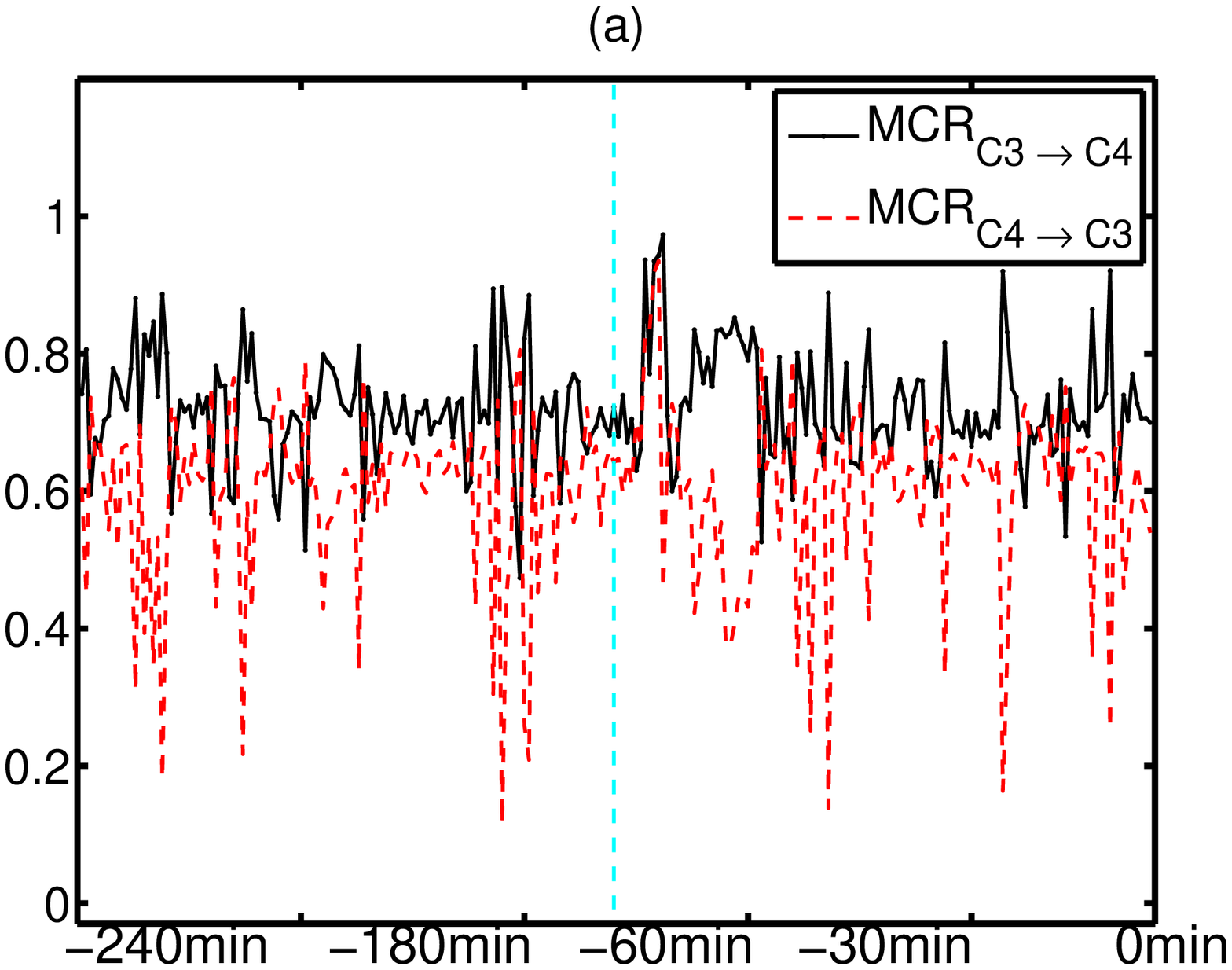}
\includegraphics[height=5.0cm,keepaspectratio]{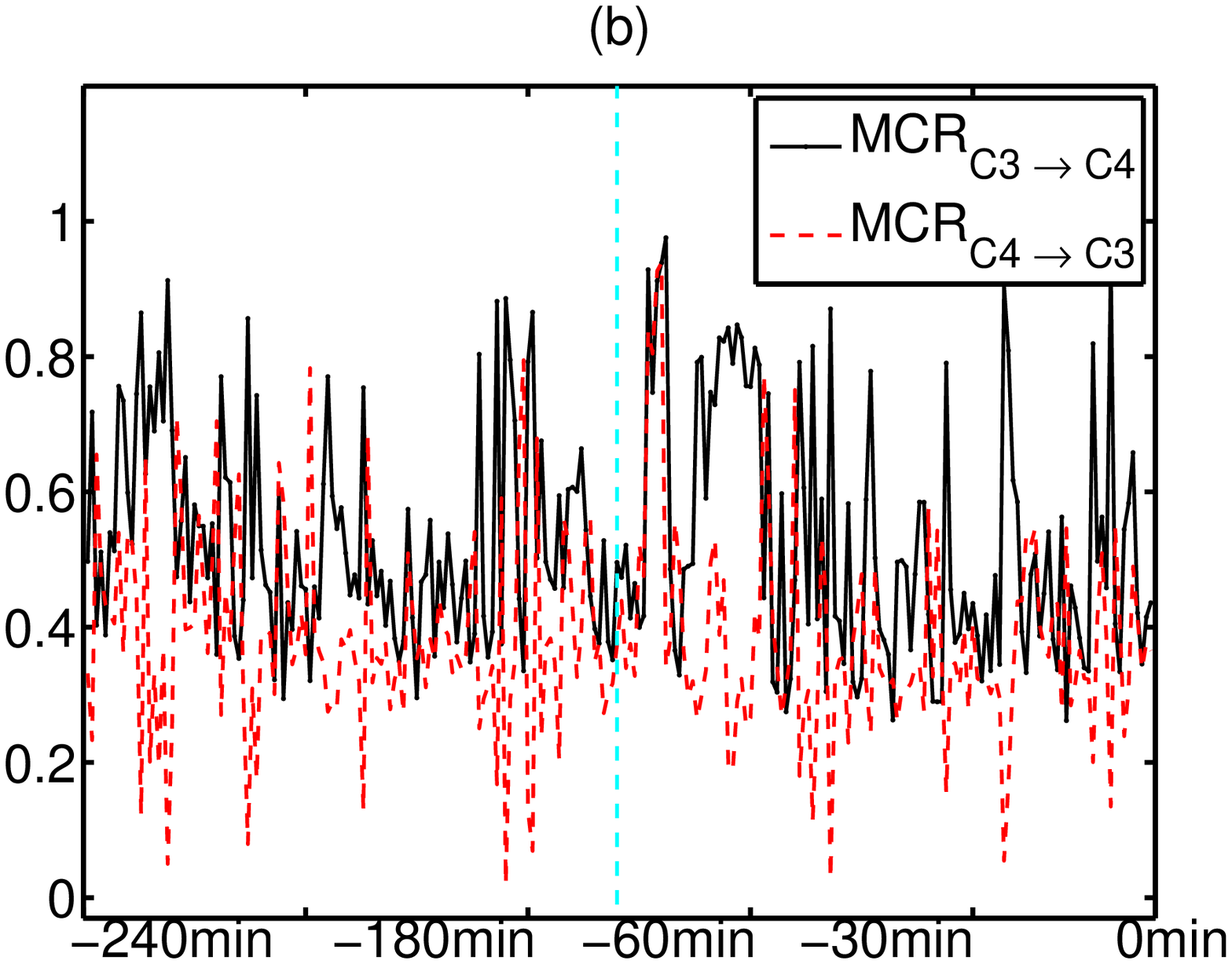}
\includegraphics[height=5.0cm,keepaspectratio]{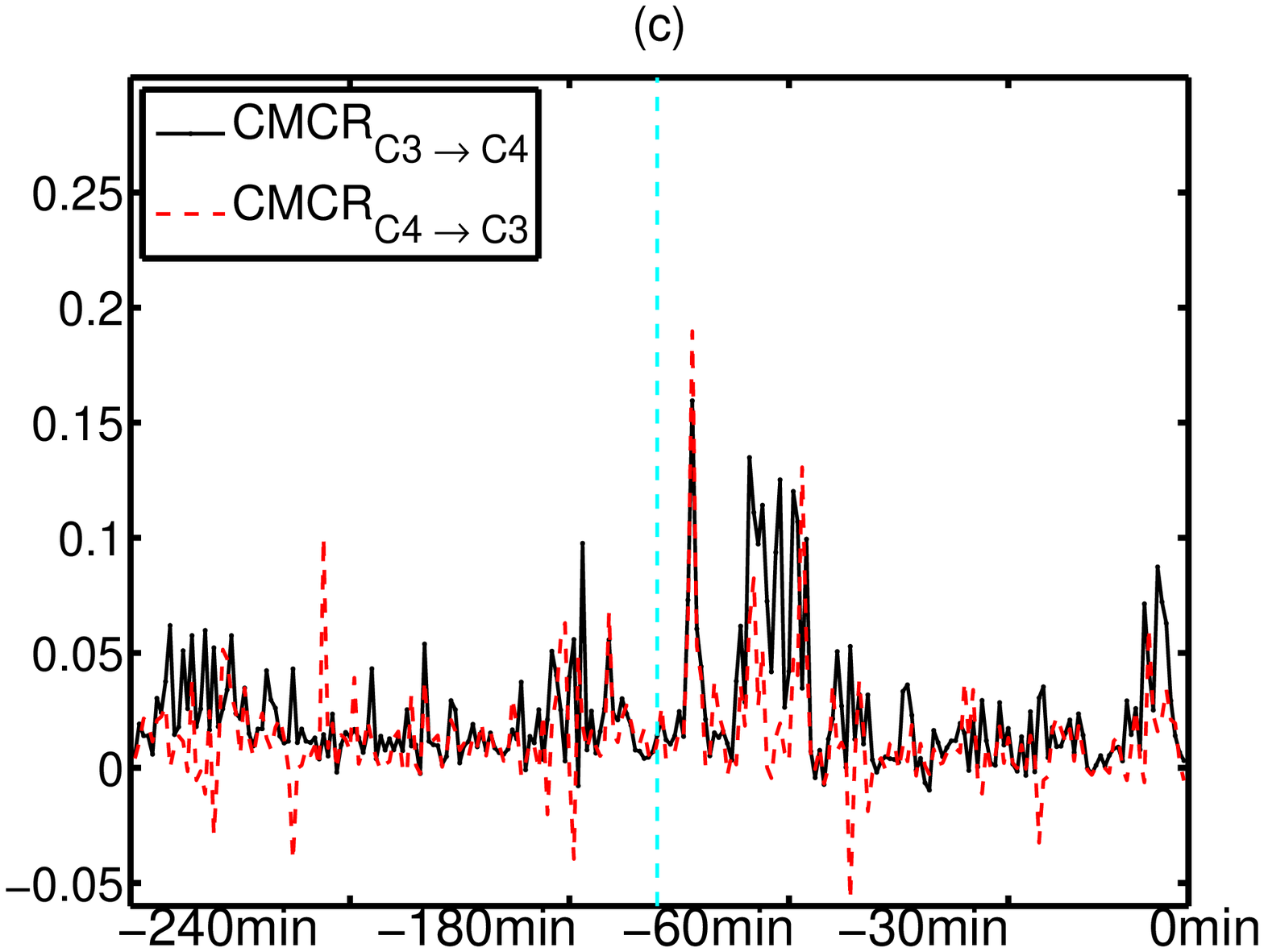}
}} \centerline{\hbox{
\includegraphics[height=5.0cm,keepaspectratio]{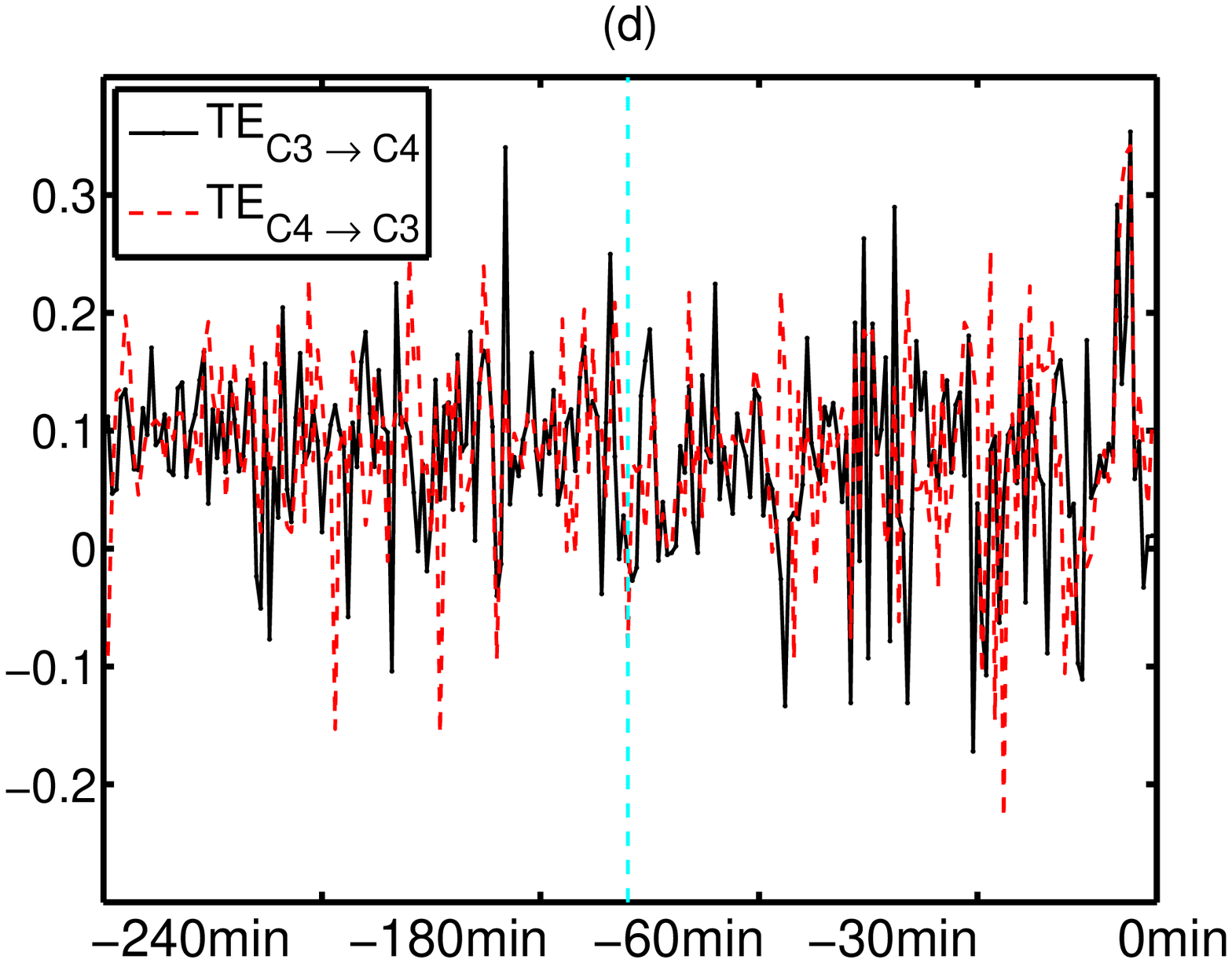}
\includegraphics[height=5.0cm,keepaspectratio]{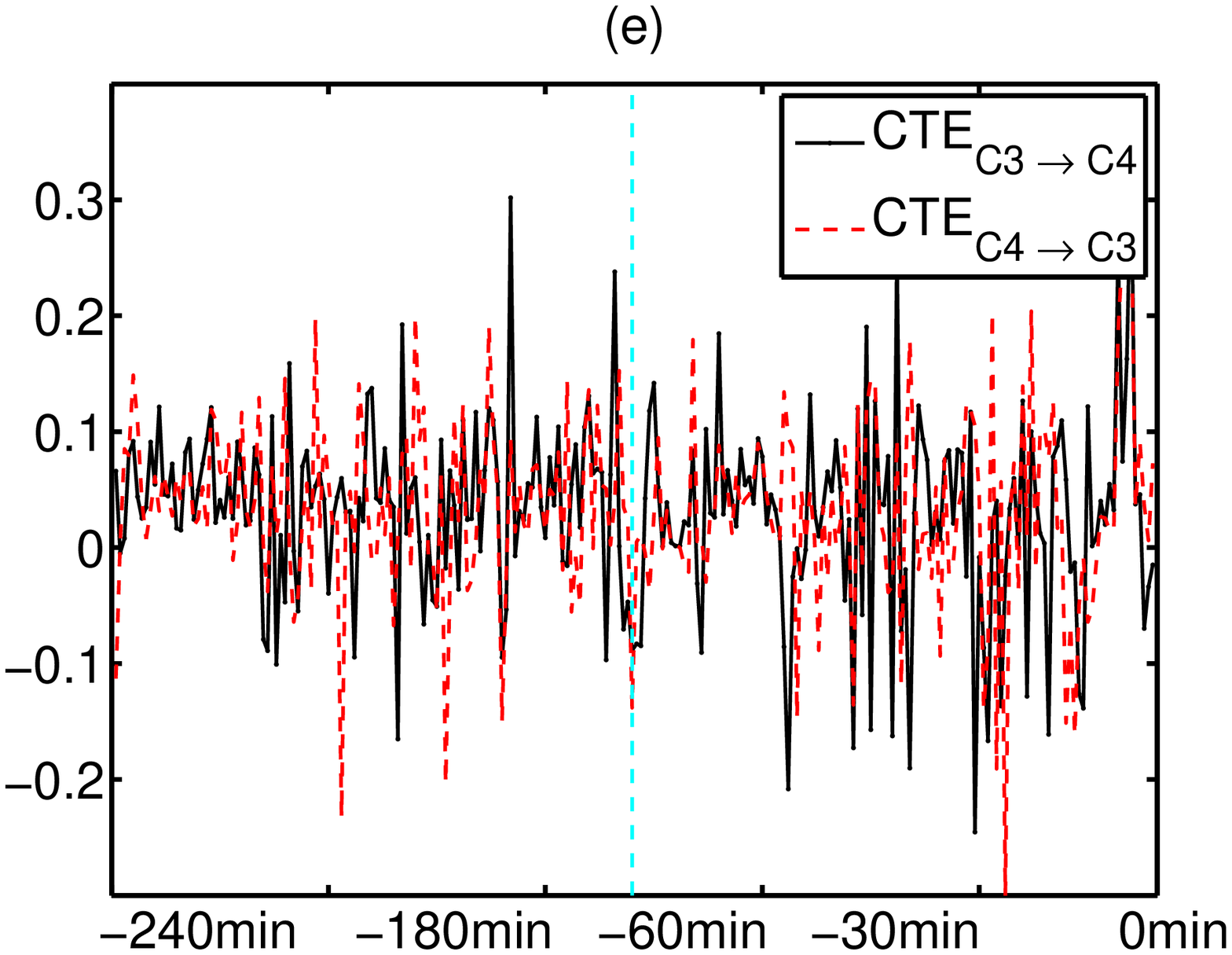}
}} \caption{(a) MCR profiles from both states (early and late
preictal) of the first seizure and both directions, for channels
as in the legend, with $m_x=m_y=3$. (b) As in (a) but for
$m_x=m_y=5$. (c), (d) and (e) are as in (a) but for CMCR, TE and
CTE, respectively. The preictal periods are indicated by the time
in min, with reference to time $0$ at seizure onset and they are
separated by a vertical dashed line.} \label{fig:gtk01}
\end{figure}

The causality measures indicate a bidirectional form of
information flow among most of the brain areas, and this is
present at both preictal states. Only few causality measures
detect a slight change in the information flow between the early
and late preictal state.
The increase of $m_x, m_y$ from 3 to 5 decreases the measure
values, as expected also from the simulation study (e.g. see
Fig.\ref{fig:gtk01}a and b for the MCR measures).
Corrected measures, as expected, give lower values than the
original measures. For the same example, CMCR shown in
Fig.\ref{fig:gtk01}c, drops to the zero level for most of the
segments regardless of the state, which shows that the interaction
observed by MCR may as well be attributed to bias in the
estimation that originates from the individual system dynamics and
state space reconstruction. A similar drop is observed for the
corrected information measures, as shown in Fig.\ref{fig:gtk01}d
and e for TE and CTE, respectively.

\begin{figure} [h!]
\centerline{\hbox{
\includegraphics[height=5.0cm,keepaspectratio]{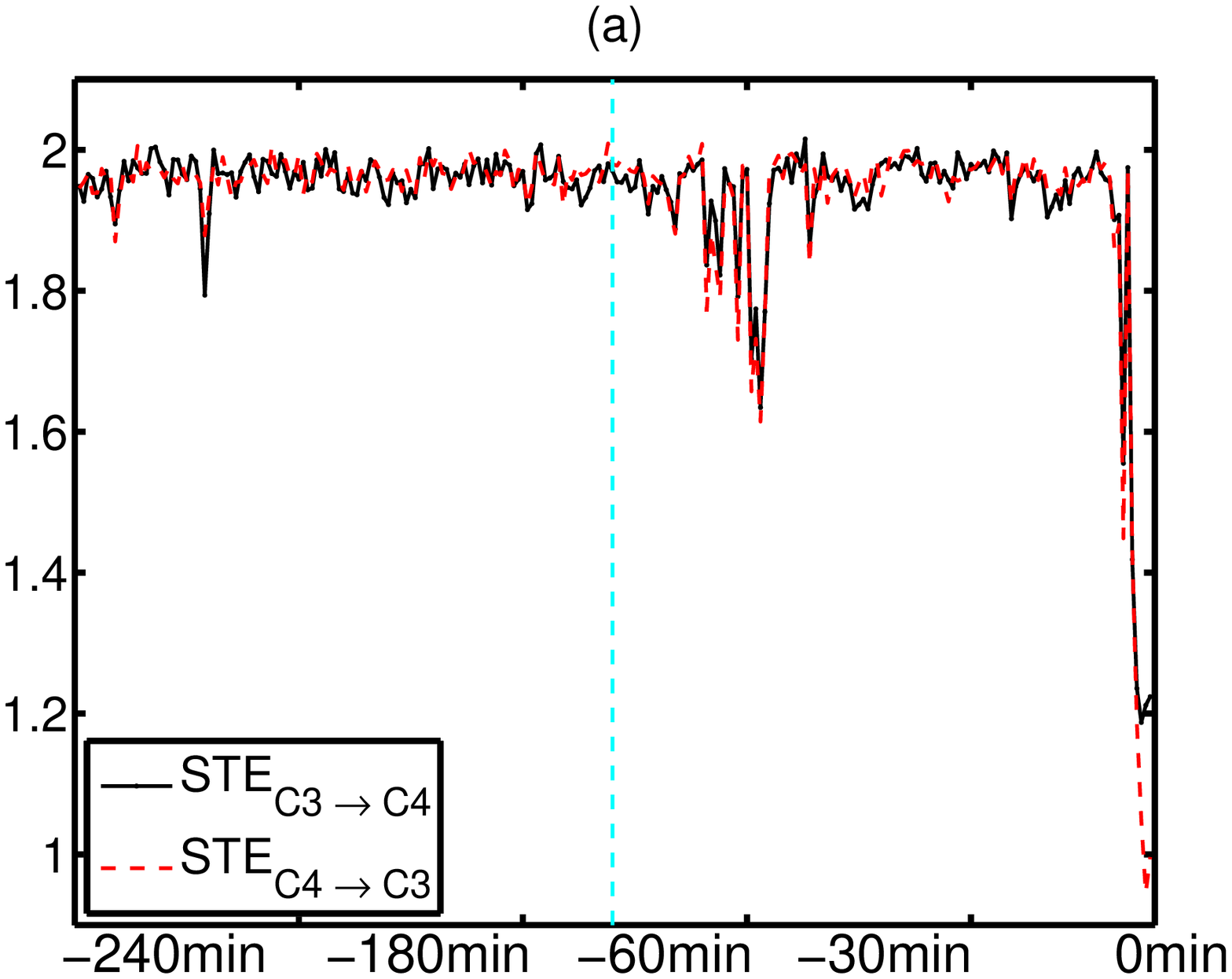}
\includegraphics[height=5.0cm,keepaspectratio]{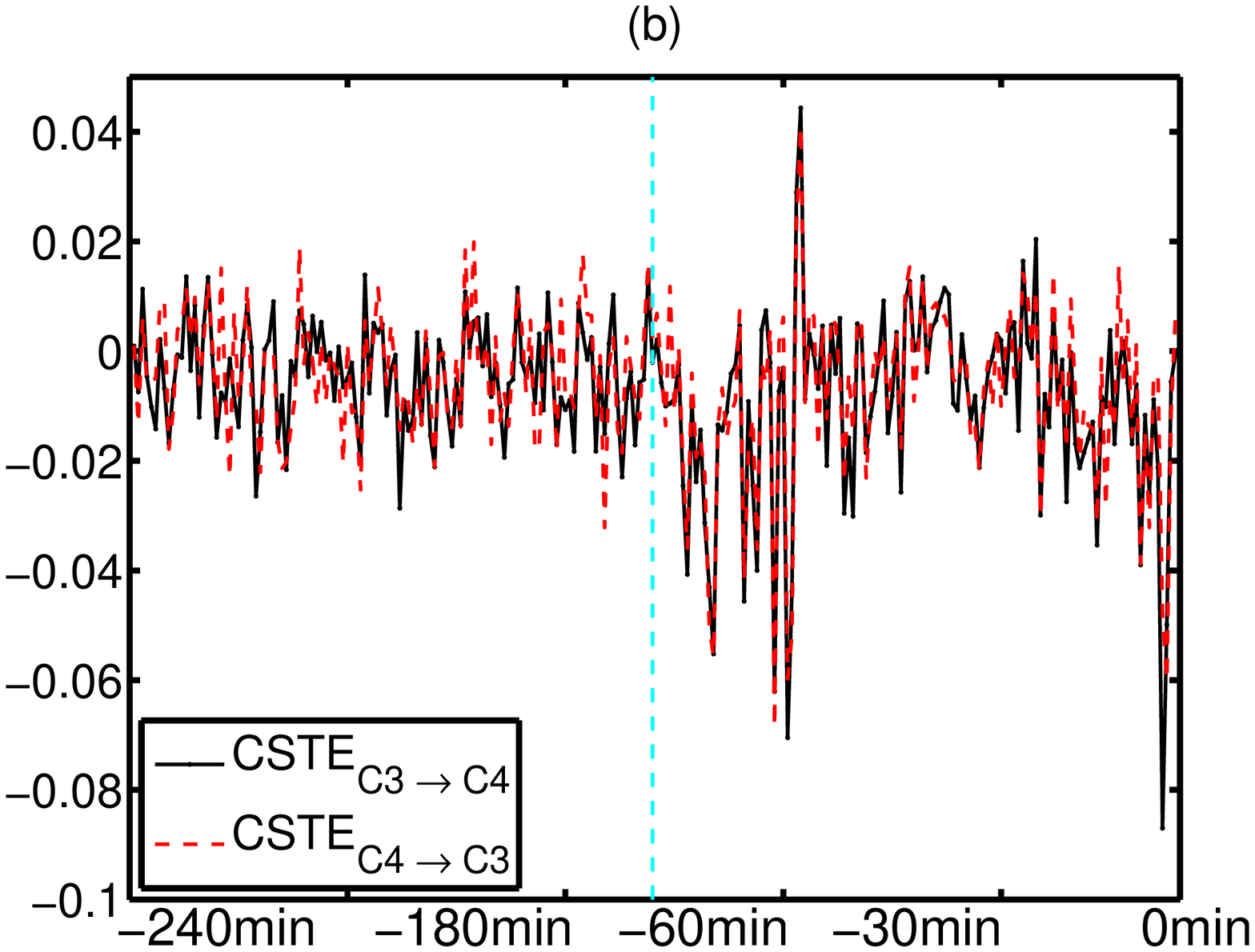}
}} \caption{(a) STE profiles from both states (early and late
preictal) of the first seizure and both directions, for channels
as in the legend, with $m_x=m_y=5$. (b) As in (a) but for CSTE.}
\label{fig:gtk01STE}
\end{figure}

There is no consistent result from all measures for the direction
of the interdependence. For example, for the pair of channels (C3,
C4) and for the first seizure, MCR measures suggest that C3 drives
C4, but not CMCR as shown in Fig.\ref{fig:gtk01}, whereas the
information measures indicate a bidirectional coupling. STE in
particular, for $m_x=m_y=5$, manifests an abrupt drop just shortly
before the seizure onset for all pairs of channels (see
Fig.\ref{fig:gtk01STE}a for channels C3, C4). CSTE renders this
drop, giving values around zero for all times (see
Fig.\ref{fig:gtk01STE}b).

For the second seizure of temporal lobe type, significant change
in the interdependence between the two preictal states could not
be observed, at least for the selected pairs of channels.
Bidirectional coupling is suggested by the original measures at
both states, whereas the corrected measures again give values
around zero. TE and CTE were rather unstable, exhibiting large
fluctuations across the successive segments of each preictal
state. Moreover, they had computational problems and they could
not always be calculated when $m_x=m_y \geq 5$ (correlation sum
contained zero terms due to lack of close neighboring points).

\begin{figure}[h!]
\centerline{\hbox{
\includegraphics[height=5.0cm,keepaspectratio]{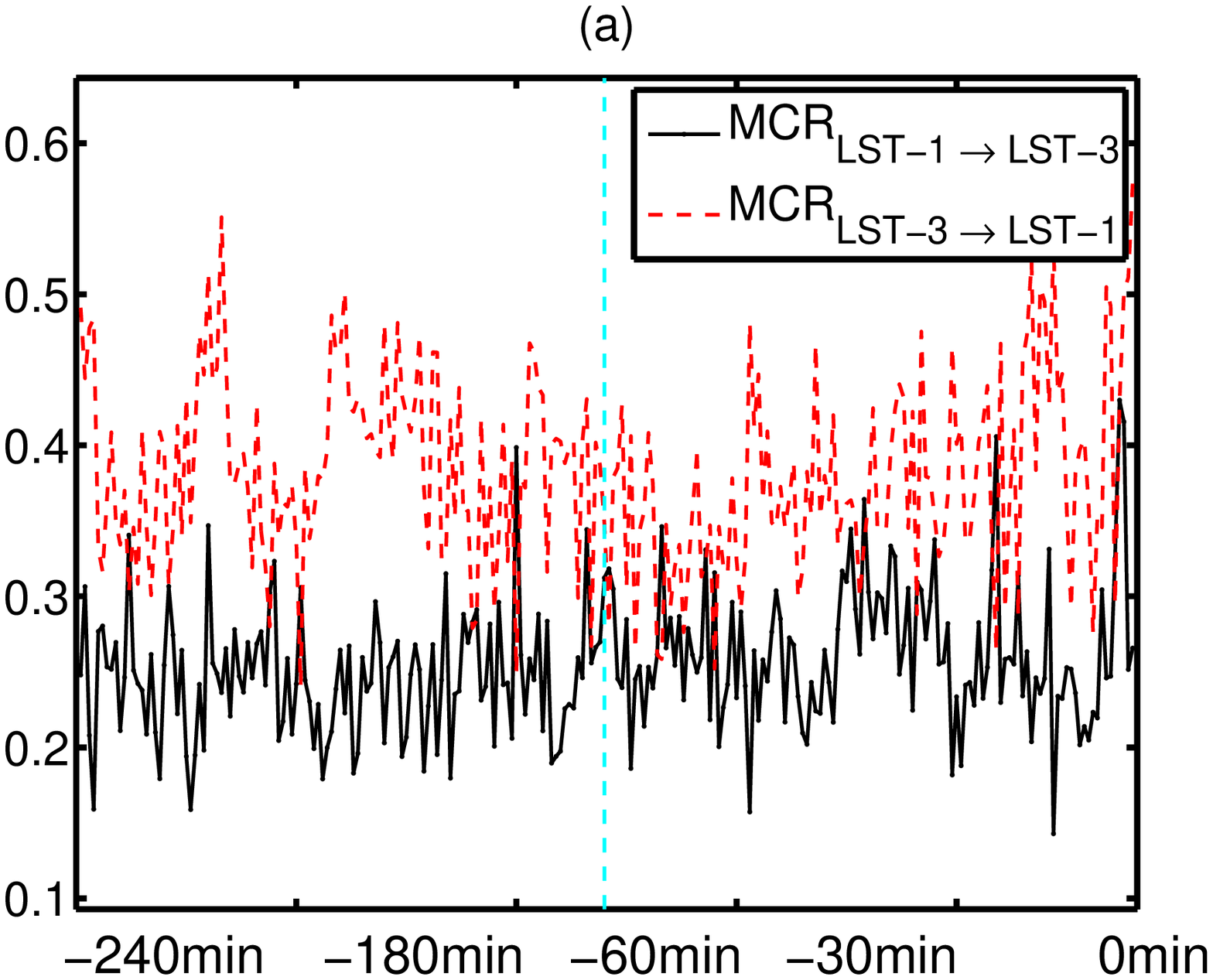}
\includegraphics[height=5.0cm,keepaspectratio]{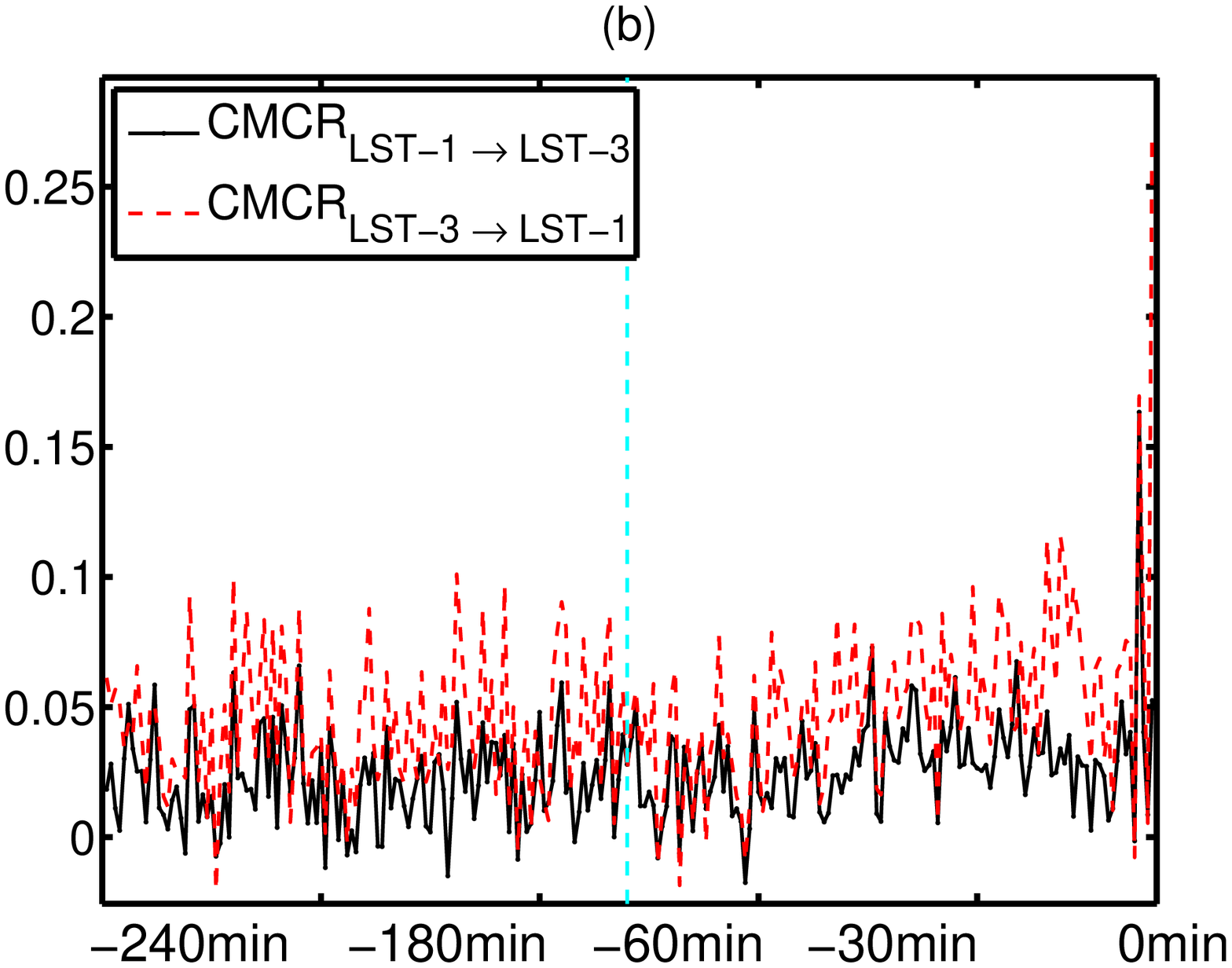}
}} \centerline{\hbox{
\includegraphics[height=5.0cm,keepaspectratio]{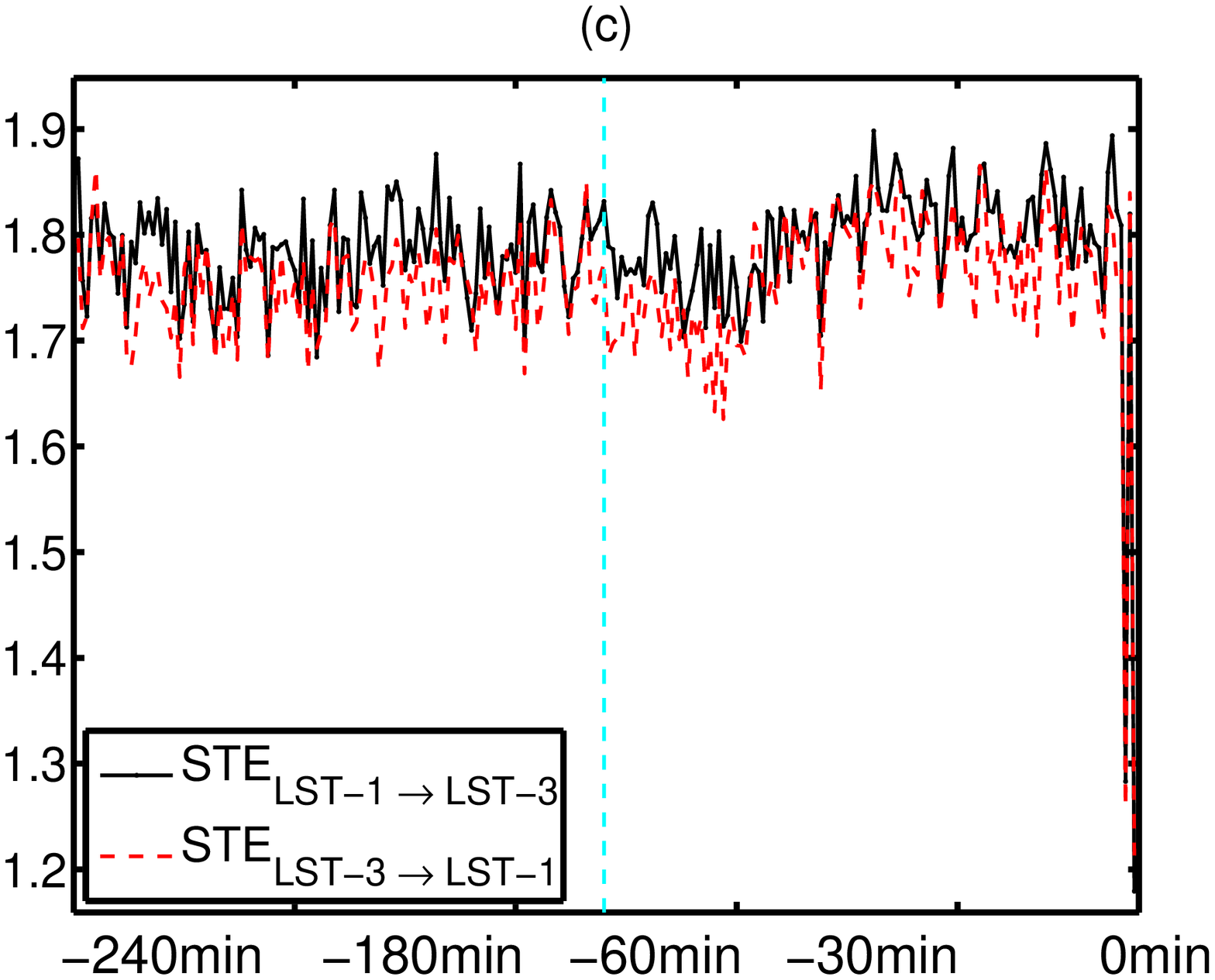}
\includegraphics[height=5.0cm,keepaspectratio]{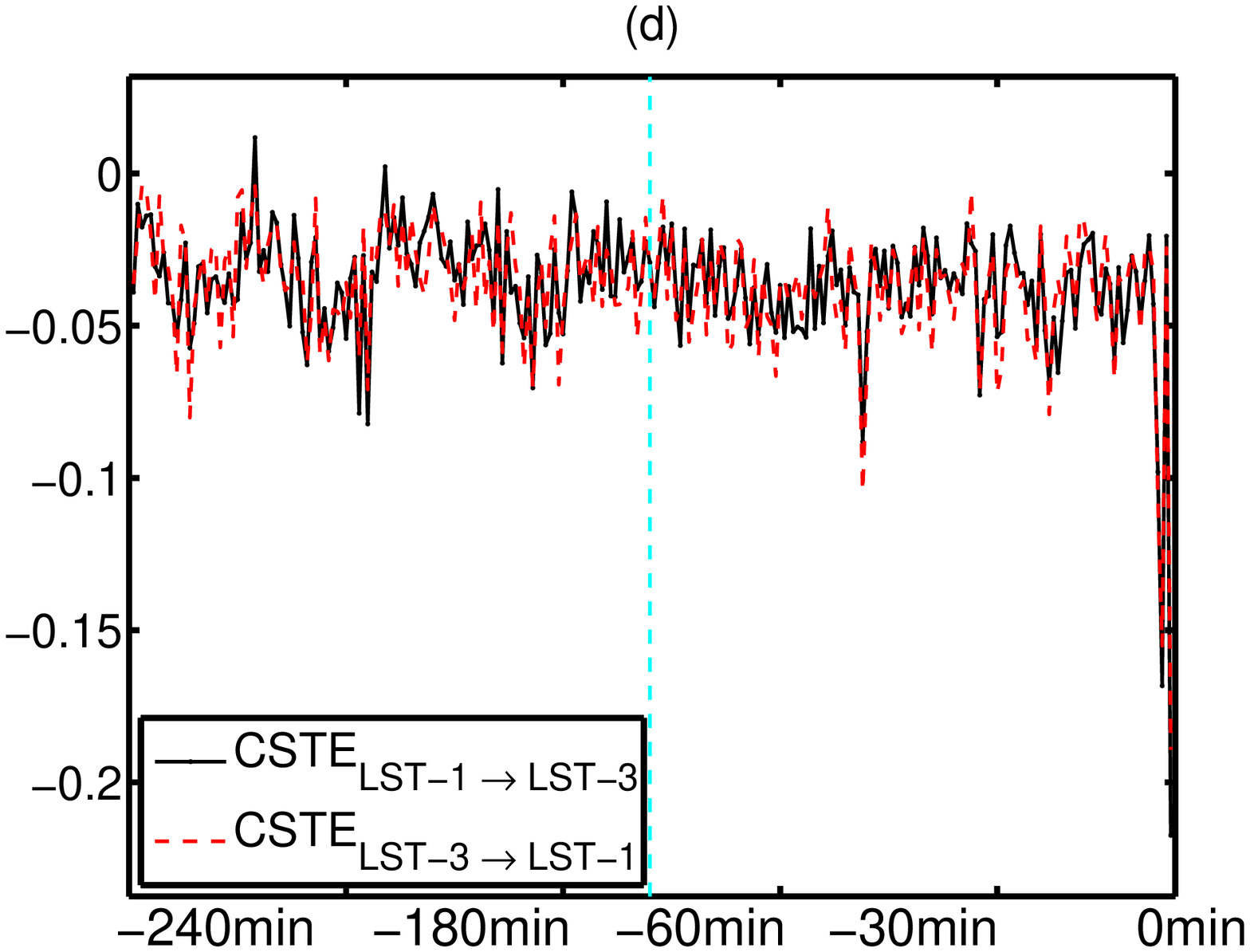}
}} \caption{(a) MCR profiles from both states (early and late
preictal) and both directions of the intracranial data, for
channels as in the legend, with $m_x=m_y=5$. (b), (c) and (d) are
as in (a) but for CMCR, STE and CSTE, respectively.}
\label{fig:Intracran}
\end{figure}

Although intracranial EEG are less noisy, no clear indication of
change in the causal effects between early and late preictal
states could be observed as well. Corrected measures again gave
values lower than the original measures (see
Fig.~\ref{fig:Intracran}), and question the coupling detected by
the original measures.

\section{Discussion}
\label{sec:Discussion}
The estimation of the strength and direction of interaction in
coupled systems from limited and possibly noisy bivariate time
series was shown to encounter a number of problems regardless of
the employed measure. It was shown that the coupling measures are
affected by a number of factors, including individual system
complexity, state space reconstruction, time series length and
noise. These factors add bias to the estimation of the strength of
coupling that may not be the same in both directions.

We concentrated on reducing the bias in each component of the
causality measures of mean conditional recurrence (MCR), transfer
entropy (TE) and symbolic transfer entropy (STE). Since there are
diverse sources of bias, we attempted to account for all of them
by assuming the value of each component measure when the systems
are not coupled. For this, we developed the idea of surrogate data
and we randomly shuffled the points of the reconstructed state
space trajectory of the driving system. Considering the
decomposition of each measure to component quantities, for each
component the average on an ensemble of realizations of the
surrogate driving trajectory was computed and subtracted from the
respective original component value. Replacing the corrected
components in the expression of the coupling measure, some of the
bias is removed. The proposed corrected measure indeed performed
as expected in simulations, but the amount of reduction of the
bias varied with the measure: for MCR the reduction with the
corrected MCR (CMCR) was small in most simulations, whereas it was
much larger in the application on epileptic EEG; for TE and STE
the reduction was larger and in most cases effective, so that the
corrected measures, CTE and CSTE respectively, were at the zero
level in the absence of coupling.

One could argue that it is intuitively more appropriate to compute
first the coupling measure on the surrogate realizations and then
take the average, instead of taking the average of the components
in the measure expression. For example, in the computation of CTE,
we take the average of the correlation sums in
Eq.\ref{eq:TECorSum} over all surrogates, while one would expect
to take the average of the whole expression for TE, which would be
equivalent to taking averages of the logarithms of the correlation
sums. The latter gives more variable estimates of TE on the
surrogates as for small values of the correlation sums we obtain
large negative logarithms. Indeed our simulations showed that this
version of CTE produces more varying results encountering also
large negative values for some realizations.

The main advantage with the corrected measures is that they
establish significance, meaning that they do not indicate
significant coupling when it is not there. This has been shown
with all tested measures, CMCR, CTE and CSTE, and for varying
conditions of system complexity (Henon maps and Mackey-Glass of
varying complexity), state space reconstruction (a range of
embedding dimensions), time series length and noise. For TE and
STE, we considered also the so-called "effective" measures,
denoted ETE and ESTE, respectively, which use a similar surrogate
approach but the random shuffling is done on the samples of the
time series. The simulation results showed that this approach
gives varying estimation of strength and direction of coupling
that often does not correspond to the real coupling. The use of
twin surrogates or time-shifted surrogates gives the same or worse
results compared to the suggested corrected measures.

The performance of the measures was also assessed by statistical
testing, where the samples for the test were the measure values on
a number of realizations. CTE and CSTE were consistently found to
be statistically insignificant in both directions in the absence
of coupling, as opposed to the original TE and STE, as well as ETE
and ESTE. In the presence of causal effect, CTE and CSTE could
identify it with the same statistical significance as TE and STE,
respectively. The correction of MCR also improved the statistical
results, but not as clearly as for the information measures.
Comparing CTE and CSTE, the simulations showed that CSTE was more
dependent on the selection of the embedding dimensions but more
robust against noise.

TE, and subsequently CTE, have computational problems when the
embedding dimension is large, at least when correlation sums are
used for their estimation, because stable statistics on
neighborhoods within a given distance cannot be established when
the state space dimension is large. This was found to be the case
for the application to EEG when the embedding dimension was larger
than 5, where TE and CTE fluctuated a lot on successive segments
of pairs of EEG channels. On the other hand, STE and CSTE were
stable and in many cases CSTE provided values close to zero,
whereas STE was always larger. CMCR also gave significantly
reduced values compared to MCR, but not at the zero level.
Interpreting these results in view of the simulation results, CSTE
was the most conservative in giving evidence for coupling, but
most reliable as well, so that when coupling was actually
indicated by CSTE it would be likely to be true coupling. We could
not find any clear evidence that there exists a particular spatial
structure of coupling at the different cortical regions we tested,
or that there is a change of the coupling structure from early
preictal to late preictal state, at least on the three EEG records
we studied. \\

{\bf Acknowledgments} \\
We would like to thank also Ralph Andrzejak for fruitful
discussions and suggestions on this work.


\end{document}